\documentclass[letterpaper,twocolumn,10pt]{article}
\usepackage{usenix-2020-09}

\usepackage[pdftex]{graphicx}
\DeclareGraphicsExtensions{.pdf,.eps,.png,.jpg,.jpeg}

\usepackage{acro}
\DeclareAcronym{pcba}{
  short=PCBA,
  long=PCB assembly,
  long-plural-form=PCB assemblies,
}
\DeclareAcronym{pcb}{
  short=PCB,
  long=printed circuit board,
}
\DeclareAcronym{ic}{
  short=IC,
  long=integrated circuit,
}
\DeclareAcronym{soc}{
	short=SoC,
	long=system on chip
}
\DeclareAcronym{bom}{
  short=BoM,
  long=bill of materials,
}
\DeclareAcronym{ipc}{
    short=IPC,
    long=Institute for Interconnecting and Packaging Electronic Circuits,
}
\DeclareAcronym{cad}{
    short=CAD,
    long=computer-aided design,
}
\DeclareAcronym{oem}{
    short=OEM,
    long=original equipment manufacturer,
}
\DeclareAcronym{aoi}{
  short=AOI,
  long=automated optical inspection,
}
\DeclareAcronym{ip}{
  short=IP,
  long=intellectual property,
}
\DeclareAcronym{cm}{
  short=CM,
  long=contract manufacturer,
}
\DeclareAcronym{ict}{
  short=ICT,
  long=in-circuit test,
}
\DeclareAcronym{fsb}{
  short=FSB,
  long=front side bus
}
\DeclareAcronym{1bl}{
  short=1BL,
  long=first-stage boot loader
}
\DeclareAcronym{2bl}{
  short=2BL,
  long=second-stage boot loader
}
\DeclareAcronym{dma}{
  short=DMA,
  long=direct memory access
}
\DeclareAcronym{smc}{
  short=SMC,
  long=System Management Controller
}
\DeclareAcronym{mmu}{
  short=MMU,
  long=Memory Management Unit
}
\DeclareAcronym{gpio}{
  short=GPIO,
  long=general purpose I/O
}
\DeclareAcronym{rgh}{
  short=RGH,
  long=Reset Glitch Hack
}
\DeclareAcronym{lpc}{
  short=LPC,
  long=low pin count
}
\DeclareAcronym{dmca}{
  short=DMCA,
  long=Digital Millennium Copyright Act
}
\DeclareAcronym{tea}{
  short=TEA,
  long=Tiny Encryption Algorithm
}
\DeclareAcronym{post}{
  short=POST,
  long=power on self-test
}
\DeclareAcronym{ldv}{
  short=LDV,
  long=lock down value
}
\DeclareAcronym{vm}{
	short=VM,
	long=virtual machine
}
\DeclareAcronym{mac}{
	short=MAC,
	long=message authentication code
}
\DeclareAcronym{bmc}{
	short=BMC,
	long=baseboard management controller
}
\DeclareAcronym{cots}{
	short=COTS,
	long=commercial off the shelf
}
\DeclareAcronym{tpm}{
	short=TPM,
	long=trusted platform module
}
\DeclareAcronym{puf}{
	short=PUF,
	long=physical unclonable function
}
\DeclareAcronym{pki}{
	short=PKI,
	long=public key infrastructure
}
\DeclareAcronym{iommu}{
	short=IOMMU,
	long=I/O memory management unit
}
\DeclareAcronym{tte}{
	short=TTE,
	long=time-triggered ethernet
}
\DeclareAcronym{pcf}{
	short=PCF,
	long=protocol control frame
}
\DeclareAcronym{pcr}{
	short=PCR,
	long=platform configuration register
}
\DeclareAcronym{drtm}{
	short=D-RTM,
	long=dynamic root of trust measurement
}
\DeclareAcronym{tcg}{
	short=TCG,
	long=Trusted Computing Group
}
\DeclareAcronym{ca}{
	short=CA,
	long=certificate authority,
        long-plural-form=certificate authorities
}
\DeclareAcronym{mitm}{
        short=MITM,
        long=man-in-the-middle,
}
\DeclareAcronym{fib}{
	short=FIB,
	long=focused ion beam
}
\DeclareAcronym{vna}{
	short=VNA,
	long=vector network analyzer
}
\DeclareAcronym{pdn}{
	short=PDN,
	long=power delivery network
}

\usepackage{array} \usepackage{multirow}

\title{SoK: A Security Architect's View of Printed Circuit Board Attacks}

\author{
  {\rm Jacob Harrison}\\University of Florida, Gainesville, FL\\Email: jacob.harrison@ufl.edu \and%
  {\rm Nathan Jessurun}\\Terraverum, Austin, TX \and%
  {\rm Mark Tehranipoor}\\University of Florida, Gainesville, FL%
}

\begin{document}

\maketitle

\begin{abstract} Many recent papers have proposed novel electrical
measurements or physical inspection technologies for defending \acp{pcb}
and \acp{pcba} against tampering. As motivation, these papers frequently
cite Bloomberg News' ``The Big Hack'', video game modchips, and
``interdiction attacks'' on IT equipment. We find this trend concerning
for two reasons. First, implementation errors and security architecture
are rarely discussed in recent \ac{pcba} security research, even though
they were the root causes of these commonly-cited attacks and most other
attacks that have occurred or been proposed by researchers. \textbf{This
suggests that the attacks may be poorly understood.} Second, if we
assume that novel countermeasures and validation methodologies are
tailored to these oft-cited attacks, then \textbf{significant recent
work has focused on attacks that can already be mitigated instead of on
open problems.}

We write this SoK to address these concerns. We explain which tampering
threats can be mitigated by \ac{pcba} security architecture. Then, we
enumerate assumptions that security architecture depends on. We compare
and contrast assurances achieved by security architecture vs. by
recently-proposed electrical or inspection-based tamper detection.
Finally, we review over fifty \ac{pcba} attacks to show how most can be
prevented by proper architecture and careful implementation.
\end{abstract}

This manuscript is an extended version of a USENIX Security 2025 paper.
Most of the extra content concerns the Xbox and Xbox~360. We believe
that additional detail in these sections is both entertaining and
valuable for PCBA security researchers. 

\section{Introduction} \label{introduction}

\begin{figure}
	\centering
	\includegraphics{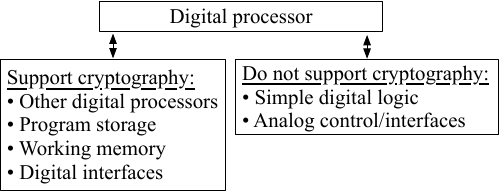}
	\caption{Only interactions where both parties can use cryptographic
	protocols can be protected by security architecture.}
	\label{fig:pcb}
\end{figure}

When a digital security architect looks at a \ac{pcba}, they see
interactions between digital processors\footnote{We use the term
``digital processor'' loosely to refer to a chip that has digital inputs
and outputs, can execute cryptographic protocols, and can condition its
execution on signature verifications and integrity checks. This could be
a general purpose CPU, a microcontroller, an FPGA, an ASIC, etc.} and
other subsystems. These interactions may be placed into two categories
as shown in Figure~\ref{fig:pcb}: those that \emph{can} be protected by
cryptography-based security architecture and those that \emph{cannot}.
Architectural defenses for interactions that support cryptography are
described in the following paragraphs.\footnote{Note that we do not
equate `security' with `cryptography'. In the words of Peter Neumann:
``if you think cryptography is the answer to your problem, then you
don’t know what your problem is''~\cite{nsa_developing_2012}. Rather,
our point is that many \ac{pcba} attacks can be prevented by a security
architecture that accounts for board-level threats.}

\textbf{Security architecture can prevent adversaries from impersonating
or replacing legitimate components, and it can prevent snooping or
tampering with inter-component communications}. When two digital
processors interact, they can authenticate each other using \ac{pki} or
shared secrets, then use cryptographic protocols to set up an encrypted
and integrity-protected channel. Adversaries cannot impersonate or
substitute components unless they can learn real authentication keys.
Similarly, a secure channel between components protects data
confidentiality as it crosses the \ac{pcb} and enables processors to
detect tampered communications based on secrets that a board-level
adversary will not know.

\textbf{Snooping and modification of data stored in non-volatile or
working memory can be prevented.} When a processor reads a program or
data from non-volatile storage (e.g., EEPROM, flash, or a hard disk), it
can verify the authenticity and integrity of the code/data using a
signature (HMAC, RSA, ECC, etc.). This prevents \ac{pcba} attacks that
hijack the processor by modifying stored code or data. Similarly,
signatures can prevent \ac{pcba} attacks that corrupt code or data in
working memory (e.g., DRAM). To prevent snooping, data can always be
encrypted before it leaves a processor and decrypted when it gets read
back.

\textbf{Peripherals can be authenticated, communications with them can
be secured, and sound design can protect a system from malicious
peripherals.} When a processor interacts with peripheral devices (e.g.,
via USB or PCIe) or with a remote computer, it can use the same
techniques described in the preceding paragraphs to either
(a)~authenticate and encrypt/integrity protect communication with the
device, or (b)~ensure that data written or read from the peripheral
stays secret and is not corrupted. If requirements dictate that a
processor must interact with arbitrary (untrusted) devices, the system
can be designed to limit peripherals' access to critical resources,
e.g., using an IOMMU.

On the other hand, \textbf{the digital architect cannot protect their
processor's interactions with simple digital logic} (e.g., glue logic,
push-buttons) \textbf{or analog circuits} (e.g., analog control, analog
transducers, power regulators). This means that \ac{pcba} adversaries
can potentially use these unprotected inputs to manipulate a processor,
and they can prevent a processor from properly controlling analog or
simple digital peripherals. If the architect is lucky, it will be
possible to use heuristics or human intervention to prevent unprotected
inputs from manipulating a system into doing bad things.

\section{Systematization of PCBA tamper defenses}
\label{sec:digital_architects_view}

This section explains \ac{pcba} security architecture's foundational
assumptions, then explores when assurances based on security
architecture complement, and are redundant with, those achieved by
electrical and optical tamper detection.

\subsection{PCBA security architecture assumptions} The design patterns
from Section~\ref{introduction} assume that \textbf{signals inside a
chip are safe from \ac{pcba} attacks.} We assess that this is a
reasonable assumption for most threat models, and that it is a
\emph{beneficial} assumption even in threat models where it is not true.

\subsubsection{IC vs. PCBA attacks}
\label{separating_ic_from_pcba_attacks}
Chip-level tampering is performed against the silicon or metal layers of
an \ac{ic} die. Examples include ``Trojan horse'' \acp{ic}, FIB edits,
microprobing, and photon emission microscopy. These attacks cause a
\emph{chip} to violate its functional or security requirements: it
either operates incorrectly with respect to its inputs or leaks internal
state that could not have been inferred from the untampered chip.

On the other hand, board-level attacks operate on a \ac{pcba}'s
components, traces, or substrate. Examples include component addition,
removal, or substitution, as well as changes to substrate or routing.
Board-level adversaries are limited to manipulating or monitoring a
chip's inputs and outputs to coax it into breaking the system's
security.

Some might question whether our chip/board boundary accommodates side
channel analysis and fault injections, which seem to infer or influence
a chip's internal state even though they are board-level attacks. We
believe it does: although these attacks exploit more arcane aspects of a
chip's operation than, e.g., changing signals on a bus, they still use
behaviors of untampered chips to break system security. In other words,
operating voltage, electromagnetic fields, reset lines, etc. are simply
more signals an adversary may manipulate or monitor. The security
architect is responsible for understanding and mitigating their system's
weaknesses to fault injection and side channels, just as with more
ordinary tampering attacks.\footnote{We assume the reader is familiar
with most of the architectural defenses we discuss, but we offer
additional explanation of how the architect may defend against fault
injection and side channels.

To mitigate fault injection, a security architect might specify that
certain data reads or conditional instructions be performed multiple
times with random delays inserted between each iteration. They might
also be mindful of processor outputs that can serve as stable timing
references before some security-critical event. If they have the luxury
of working with chip designers, they can also ask for internal filtering
on reset signals and power rails to change components' basic
vulnerability to fault injection.

Side channels are trickier to mitigate in
general~\cite{schaumont_hide_but_cant_verify}, but, as we understand the
situation, there are two classes of side channels: (1)~those that leak
keys given very few traces, and (2)~those that require traces from many
thousands of cryptographic operations. The security architect must
instruct their implementors to avoid the first class of side channels.
Completely mitigating the second class is challenging, but
susceptibility to these attacks can be reduced by designing protocols
that frequently rotate keys, or by ensuring that
adversaries cannot coax a system into performing many operations that
use long-lived keys (e.g., by using long-lived keys only in response to
a signed, replay-protected message).}

\subsubsection{On the ``hardware root of trust''}
\label{the_hardware_root_of_trust} Hardware security folklore holds that
tampering attacks undermine a system's ``hardware root of
trust''~\cite{darpa_shield, bhunia_editorial_2017}. In other words, they
undermine an assumption that certain parts of a system are secure
because they are outside an attacker's reach. We perceive that \ac{pcba}
security researchers have generally accepted that this is true, which
may explain why there has been so little attention paid to security
architecture for \ac{pcba} tamper prevention.

Security architecture needs a root of trust to enforce restrictions on,
and establish trust in, other parts of a system. In traditional
cybersecurity, which focuses on \emph{software} attacks, hardware is
said to be a root of trust because software adversaries cannot change
the metal and silicon that defines hardware behavior. On the other hand,
when studying attacks \emph{on hardware}, we must consider the
possibility that attackers could tamper with a system's physical
construction.

If \ac{pcba} tampering undermines a system's root of trust, then
architectural defenses are hollow. Happily, we assess that this is not
the case. The phrase ``hardware root of trust'' is too reductive; if
finer distinctions are made, \textbf{hardware that is difficult to
attack can serve as a root of trust against attacks on more vulnerable
hardware.} This distinction is important because, for reasons we explain
in the next subsection, \ac{pcba} tampering is a more widespread concern
than \ac{ic} tampering, so different systems may have more or less
sophisticated adversaries.

\subsubsection{Why ICs are a meaningful trust boundary}
\label{sec:difficulty_of_attacking_chips}

Attacking \acp{pcba} is significantly cheaper and easier than attacking
\acp{ic}.

It is \emph{cheaper} because the features an attacker must manipulate on
a \ac{pcba} are a hundred to a million times larger than on a \ac{ic}.
As a result, \ac{ic} tampering tools (e.g., microprobing stations, FIBs,
PHEMOS) are more sophisticated to build, more costly to procure and
operate, and are therefore less accessible than those required for
\ac{pcba} attacks (e.g., JTAG emulators, soldering irons, and
stereoscopes).

It is \emph{easier} because signals of interest on a \ac{pcba} are at a
higher level of abstraction than on \acp{ic}. Whereas \acp{pcba} are
composed of tens to thousands of \emph{components}, \acp{ic} may contain
more than a billion \emph{transistors}. The reverse engineering required
to understand which transistors must be modified to backdoor a chip
dwarfs that required to understand which signals on a \ac{pcba} can be
compromised.

Considering the difficulty of attacking \acp{ic}, many security
architects make a risk assessment that an adversary's benefit from a
successful \ac{ic} attack would not justify the attack's cost and
complexity (or, equivalently, that the harm from system compromise does
not justify the expense and complexity of implementing defenses against
\ac{ic} attacks). On this basis, they exclude \ac{ic} tampering from
their threat model.\footnote{For an example, we look to video game
modchips, which have been widely cited in \ac{pcb} security research:
Tony Chen says that Microsoft was not worried about \ac{ic} tampering
when designing the Xbox One, but that ``every exposed pin'' on the
\ac{pcba} was an attack surface~\cite{chen_guarding_2019}. This was
based on an assessment that \ac{ic} tampering costs more than several
video games.} \emph{Replacement} of security-critical chips remains a
concern, so \acp{ic} must be authenticated, but once a chip's identity
is confirmed it is explicitly assumed to meet the manufacturer's
functional description, quality standards, and security requirements.

For highly-critical systems with powerful (nation-state) adversaries,
\ac{ic} attacks become a realistic threat, so countermeasures against
both \ac{pcba} and \ac{ic} tampering are needed. However, even in this
case, security architects benefit from modeling \acp{ic} as roots of
trust and mitigating board-level threats via security architecture where
possible. \Ac{ic} tamper detection will be needed regardless of how
board-level threats are handled, and using security architecture to
convert potential \ac{pcba} tampering into \ac{ic} tampering will only
make an adversary's life more difficult.

\subsection{Security architecture vs. tamper detection}

At the end of the day, architectural, inspection-based, and electrical
tamper defenses are different ways to make physical attacks costly and
difficult. Security architecture accomplishes this by converting
\ac{pcba} attacks into \ac{ic} attacks, while electrical and optical
countermeasures aim to detect board-level probing and hardware
modification. These approaches can be complementary, or redundant. This
section further explores strengths and gaps in each approach.

\subsubsection{A brief review of tamper detection}
\label{review_of_tamper_detection}

Recent work has focused on two kinds of tamper detection:
\textbf{(1)~electrical characterization}, and \textbf{(2)~physical
inspection}.

\paragraph{Electrical characterization} Electrical detection senses
parameter changes caused by hardware modification or probing.

\textbf{Many approaches try to measure intrinsic electrical variations
from components and the manufacturing process that are (1)~difficult to
mimic and (2)~likely to change if a system is tampered.} Various
measurements and equipment/sensors have been proposed, such as
impedance~\cite{zhang_robust_2015, zheng_design_2017,
edwards_authenticating_2017, kumar_devfing_2021, guo_mpa_2017,
nishizawa_capacitance_2018}, resonant
frequencies~\cite{mcguire_pcb_2019}, signal
reflections~\cite{xu_bus_2020}, propagation
delay~\cite{paul_silverin_2021, oksman_method_2020, xu_runtime_2021},
and more. A recent trend is to characterize impedance over a frequency
range~\cite{safa2024parasiticcircusonfeasibilitygolden,
ImpedanceVerif_2022, 10153638, scattering_param_2023}.

\textbf{Other techniques measure dynamic parameters in search of
anomalies that could indicate tampering or counterfeit components.} For
example, \cite{cobb_intrinsic_2012}~measures a running circuit's EM
emanations, and~\cite{bergman_battelle_2016, piliposyan_hardware_2020}
measure dynamic power. Similarly, \cite{MAILLARD2023104904} proposes
monitoring a digital processor's control flow via its EM leakage.

A final class of electrical tamper detection uses an \textbf{active
enclosure} around sensitive subsystems to detect penetration
attempts~\cite{immler_b-trepid_2018}, similar to the old IBM
HSMs~\cite{anderson_security_2020}.

\paragraph{Physical inspection} Physical inspection-based tamper
detection takes images of a board and searches them for signs of
tampering. Innovations in this area have focused on novel imaging
modalities and automated inspection algorithms.

\textbf{Different imaging modalities reveal different information about
a board}. Lots of recent work has focused on analyzing images from
simple visual light cameras~\cite{10.1145/3588032}, which quickly and
cheaply reveal colors, textures, part markings and logos, silkscreen,
and other features. X-ray, both two-dimensional and computed tomography,
has been investigated for viewing \ac{pcb} connectivity, inspecting
inside components, and finding components hidden between \ac{pcb}
layers~\cite{10647049}. Other, more exotic modalities have also been
proposed, such as terahertz for material
analysis~\cite{true_review_2021}. A summary of research on additional
modalities may be found in~\cite{10.1145/3401980}.

\textbf{Images of a board are analyzed with computer vision, image
processing, and AI algorithms.} Different algorithms process information
from different modalities, but visual light and x-ray inspection have
received the lion's share of attention because these are the most mature
imaging technologies. For visual light images, algorithms for component
classification~\cite{component_detection}, board text, part number, and
logo extraction~\cite{10.31399/asm.cp.istfa2021p0012}, texture
analysis~\cite{dhanuskodi_counterfoil_2020, ghosh_recycled_2019}, pin
counting and measurement~\cite{pinpoint}, laser mark
characteristics~\cite{laser_mark_analysis}, and many others have been
proposed to assist in deciding whether a board is as expected. For X-ray
inspection, researches have focused on extracting \ac{pcb}
connectivity~\cite{9707715, 10.1145/3606948}. At least nine different
\ac{pcb} image data sets have been gathered and annotated to contribute
to machine learning model training~\cite{10.1145/3588032}, including of
x-ray images~\cite{cryptoeprint:2022/924}, and tools have been built to
accelerate and refine annotation of new
data~\cite{jessurun_component_2020}.

\paragraph{Online vs. offline} Offline techniques check for attacks at a
moment in time but do not protect a system while it is in service.
Online techniques protect a system continuously once they are installed
and activated. Offline electrical techniques use bench top instruments,
ranging from multimeters, to oscilloscopes, to \acp{vna}, whereas online
approaches incorporate sensing circuitry into a system's design. Most
inspection-based approaches are offline, but a few online approaches,
e.g., using IR to monitor for temperature changes that could indicate
runtime anomalies~\cite{10.1145/3401980}, have been proposed.

\paragraph{Golden-based vs. golden-free} Tamper detection may compare
measurements of a suspect system to a ``golden'' model, or to an earlier
measurement of the system. Often, golden models are difficult to obtain
because they require either (a)~extensive \textit{a priori} knowledge of
a system and sophisticated modeling, or (b)~a trusted sample from which
measurements can be taken. Additionally, golden-based analysis is
challenging because measurements must simultaneously be \emph{sensitive}
to changes caused by hardware manipulation and \emph{insensitive} to
changes from environmental or process variation. In electrical tamper
detection, both golden-based and golden-free techniques have been
proposed. Physical inspection usually needs golden images because
reliable measurements for golden-free assessment are difficult to
obtain.\footnote{Optical traits such as device footprints, chip texture,
marking characteristics, etc. are rarely specified in enough detail in
datasheets to provide meaningful golden-free measurements. Moreover,
allowable tolerances on each feature (pin pitch/length/spacing, package
height/width) vary significantly, making it harder to identify
suspicious components in golden-free analysis.}

\subsubsection{Reasoning about security}
\label{reasoning_about_security}

Even though architecture and tamper detection both work by increasing
attack cost and difficulty, reasoning about security from tamper
detection requires more knowledge of attacks and measurements of
countermeasure performance.

Security architecture operates on mathematical models. Systems can be
abstracted as automata, security properties can be formally specified,
and exploits are, to use the words of Shubina and Bratus, ``constructive
proofs'' that a system's architecture or implementation do not satisfy
its security requirements\cite{sergey_bratus_invited_2017}. When dealing
with such models, it is not important whether a malicious device is
large, small, or buried between the layers of a \ac{pcb}. For example, if the
wire targeted by a \ac{pcba} attack carries encrypted and authenticated
data, it will not be possible to snoop data carried over the wire or to
modify it without detection, no matter how physically subtle an attack
may be. This abstraction is enabled by the model of \acp{ic} as trust
boundaries. In sum, \textbf{architectural security is indifferent to an
attack's physical properties.}

Conversely, electrical and optical tamper detection make \emph{physical
measurements} of an attack. This means that factors like the amount of
capacitance that a probe adds to a victim circuit, or whether an attack
is implemented on the surface of a \ac{pcb} or between its layers,
materially impact countermeasure performance. More accurate tamper
detection schemes can detect more subtle attacks, thereby increasing
cost and difficulty of tampering. Also, tamper detection is affected by
uncertainties from measurements, manufacturing process variation, and
the environment.\footnote{Note that the presence of measurement
uncertainty implies security assessments based on tamper detection
should be considered \emph{statistical} problems. As tamper detection
research matures, we hope to see countermeasures evaluated in terms of
the \emph{likelihood} that they detect a particular attack.} Thus,
\textbf{to reason about security from tamper detection, we need physical
measurements of how subtle attacks can be, how sensitive a particular
tamper detection technique is, and how much uncertainty should be
expected in measurements.}

\looseness=-1
Unfortunately, we currently lack much of this information. Little
research has been conducted on how \ac{pcba} attacks can be made subtle.
We have found few works that use physical measurements to describe
disturbance caused by realistic attacks or tamper detection performance,
and we know of no comprehensive studies of environmental or
manufacturing uncertainties in tamper detection. Instead, countermeasure
performance is typically expressed in terms of concrete test cases
detected, and most countermeasures are evaluated with different test
cases. Validation using ad-hoc test cases, or with benchmarks whose
relative difficulties and relation to the general tampering threat is
unknown, make it difficult to compare different techniques' performance
or assess how effective tamper detection techniques will be against
realistic attacks.

\subsubsection{Picking appropriate defenses}
\label{picking_appropriate_defenses}

With all of this background in mind, we consider strengths and
weaknesses of tamper detection and architecture. Different defensive
strategies are appropriate for different systems and threat models, but
as a general rule of thumb:

\textbf{Start by applying the security architecture techniques from
Section~\ref{introduction} to protect a system's digital functions.}
There are several reasons that security architecture should be a
system's primary defense where it can be applied. (1)~Some secure
\ac{pcba} design patterns, such as cryptographic firmware verification
and integrity checks of critical data, will already be needed because
they protect a system against software-only/remote attacks. (2)~Because
it has been used for years to prevent software exploits, security
architecture is well-understood, widely used, and supported by
commercial hardware. (3)~The cost and difficulty that proper security
architecture incurs to attackers is easy to reason about. (4)~With a bit
of support from component vendors, architectural defenses can secure a
system through its entire lifecycle, from before a \ac{pcba} is
assembled\footnote{For example, \cite{st-sfi} is a
commercially-available implementation of component authenticity checks
for establishing initial trust in a system before imbuing it with
secrets or firmware. When the system is first provisioned (perhaps at an
untrusted facility), the processor's authenticity is established via
\ac{pki}, secrets are protected all the way down to the \ac{ic} die, and
security configuration is set atomically with firmware installation.
This establishes strong initial trust as long as \ac{ic} tampering is
not in scope, and, as discussed in
Section~\ref{sec:difficulty_of_attacking_chips}, if it \emph{is} in
scope, you're going to need chip-level defenses regardless of how you
deal with board-level threats.} to when it sits in an adversary's lab.

Next, \textbf{design tamper detection to fill security architecture gaps
or provide defense in depth.} We consider four ways that electrical or
inspection-based tamper detection can complement or augment security
architecture. (1)~Tamper detection must be used to defend analog or
simple digital functions because they do not support cryptography, as
explained in Section~\ref{introduction}. (2)~There are some attacks on
system availability that security architecture cannot prevent.
(3)~Tamper detection may enable a system to respond to sophisticated
secret extraction attacks. (4)~Defense-in-depth, where tamper detection
and security architecture are deployed redundantly, may be desired as a
hedge against implementation errors or security architecture failures.

Finally, \textbf{recognize that some tamper detection techniques only
work on certain kinds of attacks.} It is important to ensure that
implemented tamper detection techniques match threats to-be-defended.

\paragraph{System lifecycle considerations}

\begin{figure}
  \centering
  \includegraphics{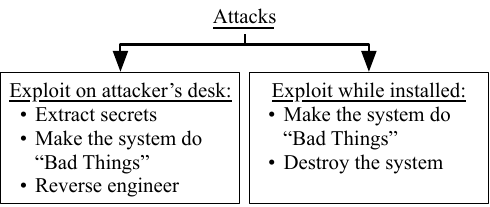}
  \caption{Depending on what attacks must be stopped and when attacks
  are anticipated, different defenses are needed.}
  \label{kinds_of_attacks}
\end{figure}

Figure~\ref{kinds_of_attacks} enumerates four goals an attacker might
have for a tampering attack. Some goals only require an attack to
succeed once, while a \ac{pcba} is in the adversary's hands. Others
require that hardware modifications persist until the \ac{pcba} is
installed and in service.

Offline tamper detection techniques can stop attacks that occur before a
system is fielded, but they cannot protect a system while it is on the
attacker's desk. To protect a system after it enters service, security
architecture or online tamper detection are required. However, even
though there are online inspection-based defenses, these can only
protect a \ac{pcba} that stays within its imaging environment.
Therefore, \textbf{to prevent attacks performed on an attacker's desk,
architecture or online electrical tamper detection are needed.}

If a system's threat model includes manufacturing-time attacks from a
board fabricator, assembler, or component vendor,
\textbf{``golden-based'' tamper detection or security architecture are
needed.} Golden-free techniques can prevent future tampering, but, if
manufacturing-time attacks are in scope, the system could be compromised
before the reference measurement or image is taken. Security
architecture is effective for checking authenticity of digital
processors and peripherals. Golden-based tamper detection must be used
to find other differences between a \ac{pcba}'s expected and actual
construction, such as authenticity checks for analog or simple digital
components, copper and substrate verification, and detecting added or
removed components on analog signal paths.

\paragraph{Attacks architecture cannot defend} As explained in
Section~\ref{introduction}, security architecture does not protect
analog control, simple digital logic, transducers, or similar circuits.
These gaps must be filled by online electrical tamper detection if
attacks can succeed in the adversary's hands, or by offline electrical
or inspection-based countermeasures if it is sufficient to check for
tampering before the system enters service.

Security architecture prevents tampering with data in
digital system interactions, but it does not prevent hardware
modification per se: \textbf{if a \ac{pcba} attack does not aim
to break confidentiality or integrity of digital data, tamper detection
may be necessary even on digital systems.} For example, consider
destructive attacks: some do tamper with code or digital
data,\footnote{Attacks like Stuxnet, and attacks that destroy flash memory
via repeated erasure, are not \ac{pcba} attacks, but they demonstrate
that control of a processor can enable physical damage to a \ac{pcba} or
a system it controls. Without proper \ac{pcba} security architecture,
similar attacks can be performed via tampering.} but there are many
creative ways to physically destroy a digital \ac{pcba} or the system it
controls that have nothing to do with code or data. A few examples have
been proposed in the literature and will be reviewed in
Section~\ref{sec:survey}. Such attacks must be detected via either
offline or online tamper detection, depending on a system's threat
model.

Finally, tamper detection may assist in preventing some kinds of reverse
engineering. Security architecture should be used to prevent adversaries
from reading firmware, contents of working memory, inter-IC
communications, and other information that reveals
how a system works, but if a system also has sensitive analog functions,
online electrical tamper detection ranging from tamper-resistant
enclosures, to photon detectors, to trace characteristic monitors are
needed. If the goal is to prevent reverse engineering the \ac{pcb}'s
hardware design, your best bet is a thick layer of hard black epoxy
imbued with dense X-ray-blocking material.

\paragraph{Reinforcing against secret extraction} To prevent attacks
that extract secret keys, sensitive data, plaintext firmware,
etc., start with security architecture and augment with tamper
detection. Low- to medium-sophistication attacks on digital assets can
be effectively foiled by using security architecture: don't let
cleartext secrets leave \ac{ic} trust boundaries, check code and data
integrity to prevent adversaries from hijacking a processor to dump
secrets, and reduce susceptibility to side channels and fault injection
through circuit design, careful firmware implementation, and key usage
policies. For the most sophisticated secret extraction attacks, online
electrical tamper detection can complement architectural defenses by
enabling the system to react to abnormal operating environments. These
sophisticated attacks may involve removing a chip from one \ac{pcba} and
placing it onto a different board that is, e.g., optimized for side
channel analysis, or for repeatedly stimulating the chip in a specific
way. Or, it could involve decapsulation, polishing backside silicon, or
other physical changes in preparation for chip-level tampering. Although
some of these attacks blur the line between \ac{pcba} and \ac{ic}
attacks, \ac{pcba} electrical defenses can nonetheless create hurdles
that increase their cost and difficulty.

\paragraph{Defense in depth} Of course, the same attack surfaces can
be protected by both tamper detection and security
architecture. One reason to implement redundant protections is as
insurance against unforeseen security architecture gaps and
implementation errors: we will see in Section~\ref{sec:xbox-360} how
such mistakes can compromise an otherwise secure system.

\section{Study of prior PCBA attacks} \label{sec:survey}

Having discussed, in abstract, how different attacks and adversaries can
be foiled by security architecture or tamper detection, we examine
actual and proposed \ac{pcba} tampering attacks to show that our
theoretical analysis also applies to real problems. We give special
emphasis to Bloomberg News' ``The Big Hack'', video game modchips, and
``interdiction'' attacks because, as we have already noted, these
attacks are commonly cited as motivation for novel tamper detection
techniques.

To structure our analysis, Table~\ref{tab:vulnerability_classes}
enumerates classes of vulnerabilities that result from failure to employ
security architecture or implement it properly. It also defines a final
category that we use for any attack that cannot be prevented with
security architecture. The Appendix summarizes our analysis and
categorizations in a table.

\begin{table}
  \centering
  \begin{tabular}{| r | p{2.9in} |}
    \hline
      & \textbf{Failure to...}\\
    \hline
    1 & Check integrity or authenticity of stored code/data.\\
    \hline
    2 & Encrypt secrets outside \ac{ic} trust boundary.\\
    \hline
    3 & Authenticate a peripheral component.\\
    \hline
    4 & Integrity-check data at \ac{ic} trust boundary.\\
    \hline
    5 & Enforce restrictions on untrustworthy peripherals.\\
    \hline
    \hline
    6 & No failure -- security architecture can't help.\\
    \hline
  \end{tabular}
  \caption{Classes of PCBA vulnerabilities caused by security
  architecture gaps or implementation failures.}
  \label{tab:vulnerability_classes}
\end{table}

\subsection{``The Big Hack''} \label{sec:big-hack-revisited}

``The Big Hack'', an article from Bloomberg
News~\cite{robertson_big_2018}, is cited as motivation by many recent
tamper detection papers~\cite{bhattacharyay_automated_2022,
botero_automated_2021,piliposyan_computer_2022,
pearce_detecting_2022,10.1145/3588032,karri_fuzzingcontrolled_2021,
piliposyan_hardware_2020,botero_hardware_2021,10153638,
true_review_2021,wang_system-level_2019,10.1145/3401980,
russ_three_2020,bhattacharyay_vipr-pcb_2022,paul_silverin_2021}.

\textbf{Limited technical details about the actual attack.} The article
alleges that a spy implant on server motherboards compromised
high-profile American companies as well as U.S. intelligence agencies.
According to the article, the implants were ``not much bigger than a
grain of rice'', they were ``gray or off-white in color'', they ``looked
more like signal conditioning couplers... than microchips'', and they
``varied in size'' between victim boards~\cite{robertson_big_2018}.
Regarding their function, the article says the chips contained only a
small amount of code and attacked the
\ac{bmc}~\cite{robertson_big_2018}, a highly privileged peripheral with
characteristically poor
security~\cite{hudson_modchips_2018,farmer_sold_2014}. The implants'
payload allegedly opened backdoors on the \ac{bmc} and retrieved exploit
code from command and control servers~\cite{robertson_big_2018}. As a
concrete example of attacks enabled by such a payload, the article
discusses how the implant might change a
password~\cite{robertson_big_2018}.

Unfortunately, these details are insufficient to analyze the implant.
The only things that the article states definitively are that the
implant hijacked control of a processor and that its payload involved
the \ac{bmc} (but not necessarily that it hijacks the \ac{bmc}). Given
these leads, researchers have prototyped plausible attacks. We study
these attacks in lieu of the actual ``Big Hack'' implant.

\textbf{No secure boot on \acp{bmc}.} Trammel Hudson proposes that, to
hijack control of a \ac{bmc}, attackers could simply replace or
reprogram the flash chip that stores the \ac{bmc}'s
firmware~\cite{hudson_modchips_2018}, as \acp{bmc} commonly load
unencrypted firmware without checking its authenticity or
integrity.\footnote{Typical \ac{bmc} firmware security may have improved
since 2018. We hope so. We have not checked.}

However, a reprogrammed flash chip could be detected if a security
auditor, like the one who allegedly found the ``Big Hack'' implant,
reads the flash and compares its contents with expected values. To demo
a more discreet, nation-state-level attack, \cite{hudson_modchips_2018}
proposes that a series resistor on the SPI data line between the
\ac{bmc} and its flash chip could be replaced with an active component
that monitors sequences of bits read and creates an open circuit at
critical moments so that the SPI line's pull-down resistors change
affected bits from `1' to `0'. Unprogrammed flash sectors read as long
strings of `1' bits, creating a blank canvas for such an implant to
inject a software payload. Assuming that the CPU reads at least a
handful of such unprogrammed bytes at the end of its firmware, all that
remains is to manipulate a branch target in the legitimate firmware to
jump to the payload bytes added in unprogrammed regions and the attacker
can hijack control of the BMC. An FPGA-based proof of concept
demonstrates the viability of this in-flight firmware
modification~\cite{hudson_modchips_2018}.

Powering a two-terminal implant parasitically off the SPI data line
would be tricky: the SPI clock would have to be recovered from strings
of alternating 1s and 0s, and the implant cannot see a chip select line
to be sure that the \ac{bmc} is reading from flash and not some other
SPI device, but Hudson argues that these challenges have known solutions
within the capabilities of sophisticated
adversaries~\cite{hudson_modchips_2018}.

Regardless of precisely how firmware gets manipulated, this
\underline{Class 1} vulnerability could be closed by industry-standard
boot security. Signature verification prevents firmware modification by
\ac{pcba} implants, no matter where they are located or how small they
are.

\textbf{Use an unprotected root shell.} Hudson notes that a far easier
way for an implant to take over a \ac{bmc} would be to connect it to the
\ac{bmc}'s serial console header, wait for the \ac{bmc} to finish
booting and print ``press enter to activate this console'', then emulate
an ``enter'' key stroke and land in a root
shell~\cite{hudson_modchips_2018}. From here, emulated keystrokes could
direct the device to change passwords and configurations in the same
manner as an administrator who has connected their server console to the
\ac{bmc}.

Monta Elkins explores this attack further. He demonstrated that a
microcontroller attached to the UART port on a commercial router could
trigger password resets and change firewall configurations to grant a
remote attacker access~\cite{monta_elkins_nation-state_2019,
greenberg_planting_2019}.

The root of this problem is that the administrator shell is not
protected by a password. This is a sane policy if physical access to a
server is deemed a sufficient barrier to defeat adversaries, but if
\ac{pcba} attacks are in scope, trusting a peripheral simply because it
is physically connected constitutes a \underline{Class 5} vulnerability.

\subsection{Video game console modchips}

Many recent
papers~\cite{paul_silverin_2021,oksman_method_2020,paley_active_2016,
bhattacharyay_automated_2022,guo_eop_2019,10.1145/3588032,
karri_fuzzingcontrolled_2021,piliposyan_hardware_2020,guo_mpa_2017,
zhang_database-free_2021,zhu_pcbench_2021} have cited `modchips' as
motivation for novel \ac{pcba} defenses. However, to the best of our
knowledge, none have analyzed any \emph{specific} modchip attacks. There
were many~\cite{fail0verflow_console_nodate,chen_guarding_2019,
andrew_bunnie_huang_hacking_2003}. Analyzing specific attacks reveals
that they were mostly caused by design and implementation mistakes in
the first two console generations. \textbf{The latest consoles with
strong, properly-implemented security architecture have no known modchip
vulnerabilities.}

Video game consoles must prevent piracy and
cheating~\cite{chen_guarding_2019}. Anti-piracy is important because
game creators like to get paid, and because console manufacturers sell
hardware below cost to entice customers while recovering profits through
game sale royalties. Anti-cheat ensures that games are fun and fair so
that customers will keep buying games and subscriptions for competitive
online play. Generally, meeting these goals requires that Microsoft
maintains control of 1)~the Xbox's processor and 2)~the foundational
secrets in the console's security architecture.

\subsubsection{Original Xbox modchips} Failures of the original Xbox's
security system were enabled by implementation mistakes and
poorly-calculated cost/security tradeoffs. Control of the processor and
cryptographic secrets were both lost repeatedly through various
mechanisms.

\textbf{First-stage boot loader mistakes.} The original Xbox's \ac{1bl}
and the key for decrypting its \ac{2bl} were \emph{transmitted in the
clear across the \ac{pcb}} from the South Bridge die, where they were
stored in a ``secret'' boot ROM, to the North Bridge, and finally over
the \ac{fsb} to the CPU where the \ac{1bl} was
executed~\cite{huang_keeping_2002, andrew_bunnie_huang_hacking_2003}.
Cleartext transmission of cryptographic secrets across the \ac{pcb} is
an obvious \underline{Class~2} vulnerability, but it was viewed as a
good cost/security compromise\footnote{Custom CPUs with sufficient ROM
for the \ac{1bl} would have been more expensive than adding a ROM to the
South Bridge~\cite{steil_17_2005}.} because Microsoft assumed the
relevant buses were too fast to sniff with amateur
equipment~\cite{steil_17_2005} and that nobody with advanced tools would
want to dump the Xbox's boot ROM. This was a costly miscalculation. The
Xbox's \ac{1bl}, including its \ac{2bl} encryption keys, were dumped and
published~\cite{huang_keeping_2002, andrew_bunnie_huang_hacking_2003}.

Then, hackers studied the \ac{1bl} and discovered that an implementation
error rendered its \ac{2bl} integrity check ineffective, a
\underline{Class~1} vulnerability. Microsoft had originally intended to
use a cipher that fed decrypted data back into the keystream such that
ciphertext modifications cause everything after the point of
modification to decrypt as garbage~\cite{steil_17_2005}. A cipher with
this characteristic enabled Microsoft to avoid including a hash
algorithm in their size-constrained \ac{1bl} by comparing the last word
of memory with a known 32-bit constant: if the last decrypted word
matched a magic value, it was safe to assume that the \ac{2bl} was
correct. However, Microsoft switched to a different encryption algorithm
that did not have this property late in the Xbox's design and did not
realize what they had lost in the change. With the new algorithm, the
last 32 bits were unaffected by changes to any of the preceding
ciphertext, enabling hackers to arbitrarily change the \ac{2bl} and, as
long as they did not change the last 32 bits, the \ac{1bl}'s integrity
check would pass.

The combination of an impotent integrity check and a leaked \ac{2bl}
encryption key enabled modchips that replaced the \ac{2bl}, either by
modifying it on the fly as the CPU fetched it or by
replacing/overwriting the flash chip~\cite{steil_17_2005}. The first
modchips disabled Microsoft's flash chip and soldered another on in
parallel, but a simpler installation was soon achieved via a
manufacturing feature that remained active: if the \ac{1bl} found no
\ac{2bl} in flash (or, if an adversary grounded the data pin to the
original flash chip to simulate this condition), the \ac{1bl} would
execute from a memory on the \ac{lpc} bus~\cite{copetti_xbox_2020,
steil_deconstructing_2006}.\footnote{This \ac{lpc} programming feature
was another cost-saving measure~\cite{steil_17_2005}. Ordering
pre-programmed memory was more expensive than ordering blank memory and
programming it in-system via the \ac{lpc} header.} Attaching an ROM to
the LPC bus required significantly less soldering than a parallel flash
chip; some were even designed with pogo pins for a solder-free
assembly~\cite{steil_17_2005, andrew_bunnie_huang_hacking_2003}. Thus,
not only were modchips possible, little technical expertise was required
to install them.

\textbf{A second failure to implement hashing.} Microsoft responded to
their console's first compromise by 1) updating the \ac{2bl} encryption
key and 2) attempting to fix their broken \ac{2bl} authenticity check.
Both measures were failures.

Updating the \ac{2bl} encryption key was ineffective because the Xbox's
hardware architecture did not change, so the new key could be dumped in
the same manner as the first (or by anybody who managed to read the
\ac{1bl}). This remains a \underline{Class 2} vulnerability.

Fixing \ac{2bl} verification \emph{should} have prevented \ac{2bl}
modification, but Microsoft made an exceptionally poor choice of hashing
algorithm. They used the \ac{tea},\footnote{The choice of \ac{tea} was
probably due to the very limited space remaining in the 512B boot
ROM~\cite{steil_deconstructing_2006}. We can speculate that without this
cost constraint, a more widely-used and secure hash like SHA1 may have
been chosen.} which yields the same digest for multiple inputs if pairs
of input bits are manipulated in a specific
way~\cite{andrew_bunnie_huang_hacking_2003, steil_17_2005}. Hackers
performed such a manipulation to change a \ac{2bl} branch target,
thereby redirecting the CPU into attacker-controlled flash. This remains
a \underline{Class 1} vulnerability.

Some hackers did not want to release exploits that depended on
Microsoft's \ac{2bl} encryption key for fear of violating the Digital
Millennium Copyright Act. They found the following other \ac{pcba}
vulnerabilities that did not need this key.

\textbf{Failure to verify that an important CPU fault actually happens.}
The ``Visor Vulnerability'', came from an incorrect assumption about the
CPU's behavior: Microsoft thought a CPU fault would occur if the
instruction pointer rolled over from \texttt{0xFFFF\_FFFF} (the highest
address in the 32-bit system) to \texttt{0x0}. They relied on this
behavior to stop the CPU and hide the secret boot ROM's contents in case
of failed \ac{2bl} authentication. In fact, a hacker tried it and
discovered that execution would happily continue at
\texttt{0x0}~\cite{andrew_bunnie_huang_hacking_2003, steil_17_2005}.
Thus, by overwriting unprotected flash data or wiring up a modchip,
attackers could control the Xbox's CPU, dump the \ac{1bl}, and recover
the \ac{2bl} encryption keys~\cite{andrew_bunnie_huang_hacking_2003}.
The Visor Vulnerability is \underline{Class 1} because it was an
implementation error that broke stored program verification.

\textbf{``Jam code'' interpreter vulnerabilities.} Another attack, the
``MIST premature unmapping attack'', was enabled by overwriting or
interposing on the same flash data as the Visor Vulnerability.

The Xbox \ac{1bl} needed to stuff several complex functions, including
RAM initialization, \ac{2bl} decryption and integrity checks, and more,
into a mere 512 bytes. To help save space, several of these operations
were performed by a virtual machine in the \ac{1bl} that executed a
sequence of ``jam codes'' stored in flash~\cite{steil_17_2005}.

The jam codes were stored unencrypted and the \ac{1bl} did not
authenticate them. To mitigate this obvious \underline{Class 1}
vulnerability, the Xbox's engineers tried to make the \ac{1bl} virtual
machine incapable of doing bad things by blacklisting jam code byte
patterns. However, their blacklist did not account for the flash chip's
address aliasing and attackers exploited this oversight to more easily
dump the \ac{1bl}~\cite{steil_17_2005}. Attackers could also sidestep
the blacklists by assembling malicious commands
byte-by-byte~\cite{steil_17_2005}; this was exploited to disable the
secret ROM while the \ac{1bl} was running, causing execution to continue
in attacker-controlled flash. The consequences for the Xbox were lost
secrets and loss of control of the
CPU~\cite{andrew_bunnie_huang_hacking_2003, steil_17_2005}.

\textbf{Vulnerabilities from a forgotten legacy behavior.} Yet another
attack leveraged a legacy CPU behavior that Microsoft's engineers had
overlooked. In brief: grounding the \texttt{A20\#} CPU pin (``the A20
gate'') caused the 20th bit of whatever address was requested by the CPU
to be set to zero. This was useful because grounding the pin changed the
Xbox's boot vector to attacker-controlled flash instead of its secret
ROM~\cite{steil_17_2005, steil_deconstructing_2006}. Once more,
overwriting flash or adding a modchip gave attackers control of the
system. This ``A20 Bug'' was also useful for dumping the \ac{1bl}.

The A20 Bug allowed attackers to sidestep stored program verification.
It is yet another \underline{Class 1} vulnerability.

\subsubsection{Xbox 360 modchips} \label{sec:xbox-360} The Xbox 360's
security architecture was greatly improved compared to the original
Xbox, but software vulnerabilities, ineffective glitching
countermeasures, timing side-channels, and various architectural and
implementation mistakes related to the disc drive enabled \ac{pcba}
attacks against the console. However, \emph{the final Xbox 360
motherboard revisions, which corrected many implementation and
architectural failures, were not vulnerable to any \ac{pcba} attacks.}

\begin{figure*}
  \centering
  \includegraphics{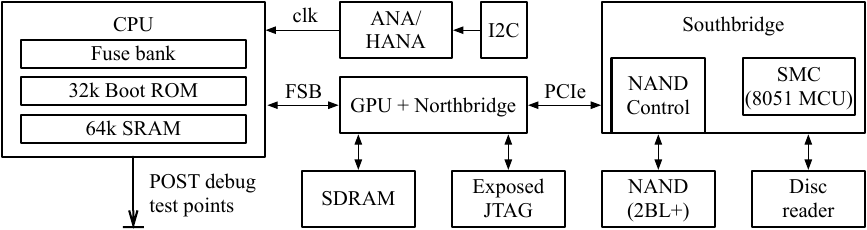}
  \caption{The Xbox 360 architecture. Visualizes key components and
  establishes terminology used in our discussion.}
  \label{fig:xbox360-architecture}
\end{figure*}

\begin{figure*}
  \centering
  \includegraphics{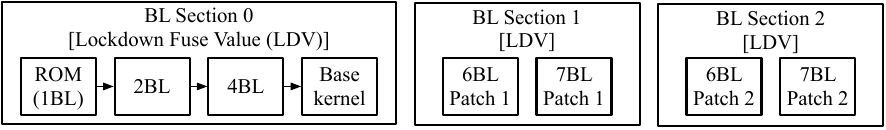}
  \caption{Summary of the Xbox 360 boot process. The \ac{1bl}, located in
  the CPU die, decrypts the \ac{2bl} into CPU-internal SRAM and verifies
  its RSA signature~\cite{free60_boot_nodate} before loading it. The
  \ac{2bl} checks the device's fuse set and may abort boot if the fuses
  indicate that the running \ac{2bl} is outdated. The \ac{2bl} also
  initializes RAM and the CPU's transparent encryption engine before
  decrypting the 4BL into RAM. Note that the the \ac{2bl} was eventually
  divided into two parts, 2BL-A and 2BL-B, but this is immaterial for our
  discussion. The 4BL verifies and decrypts the base kernel into RAM. Note
  that the manner of verification and keys involved in 4BL decryption
  changed across versions and these changes were significant for
  attackers~\cite{noauthor_jtag_nodate}. The 4BL also performs a second
  fuse check and compares the fuse set with the \ac{ldv} from the header
  of BL Section 0. If the fuses and BL Section 0 header do not match, the
  4BL launches the 6BL and 7BL from additional BL sections until one of
  their \acp{ldv} matches the device's fuse set. An update's 6BL and 7BL
  work together to patch the base kernel that was loaded into RAM by the
  4BL. Finally, after all updates have been applied, the Xbox runs the
  patched kernel from RAM.}
  \label{fig:xbox360-boot-process}
\end{figure*}

\textbf{Technical background on Xbox 360.} To aid in understanding the
attacks discussed in this section, Figure~\ref{fig:xbox360-architecture}
illustrates and explains critical components on the Xbox 360
\ac{pcba}~\cite{copetti-xbox360}, and
Figure~\ref{fig:xbox360-boot-process} illustrates the 360's boot
process.

The Xbox~360 CPU included an on-die ROM that stores the \ac{1bl}. Moving
the \ac{1bl} into the CPU die prevents bus-snooping/tampering attacks
and establishes a strong first link in the device's chain of trust. The
on-die ROM is larger than the original Xbox's boot ROM, enabling it to
store strong cryptography, and it contains more on-die SRAM than the
original Xbox as scratch space for security critical computations before
main RAM is initialized. Finally, the CPU was custom and did not inherit
compromising legacy behaviors like the original Xbox's Intel chipset.

Fuses in the CPU permanently disable its debug infrastructure and store a
unique-per-console key. Additionally, the fuses prevent firmware
downgrades and inform the boot process how many update patches are
installed~\cite{free60_fusesets}.

The \ac{smc}, which is simply an 8051 microcontroller, coordinates
various I/O related functions. It can read from flash and initiate DMA
transactions~\cite{noauthor_jtag_nodate}. There is a NAND flash below
the \ac{smc} that holds~\cite{free60_nand}: 1)~the \ac{2bl} and later
boot stages, as well as the kernel and hypervisor (all are encrypted,
and the 2BL is RSA-signed by Microsoft), and 2)~the \ac{smc} firmware
and configuration (these are encrypted, but not
signed~\cite{15432_protecting_2020_part1}).

The GPU has an exposed JTAG header. The CPU disables GPU JTAG early in
boot, but it is temporarily active~\cite{noauthor_jtag_nodate}.

The disc reader is a \ac{cots} component with custom firmware. Microsoft
sourced disc drives from various vendors but all were interoperable from
the CPU's perspective~\cite{copetti-xbox360}. Every disc drive was
paired to its CPU by a shared secret and anti-piracy data exchanges
between the CPU and disc drive were encrypted. Normal data (e.g., game
data) was not encrypted.

The Xbox 360 software security architecture is founded on a hypervisor
whose main role is to configure the CPU's \ac{mmu}. The hypervisor
ensures no memory page is both writable and executable, and it validates
program signatures before making code pages
executable~\cite{copetti-xbox360}. Different protections were applied
to hypervisor vs. non-hypervisor memory:
\begin{itemize}
  \item As the hypervisor is clearly security-critical, its code and data are
  both encrypted and integrity-protected when stored outside the CPU die
  (in DRAM). \item Non-hypervisor code pages are encrypted off-chip to
  prevent \ac{dma} attacks that hijack the processor. \item
  Non-hypervisor data pages are not encrypted because devices other than
  the CPU, which are oblivious to the CPU's encryption, must be able to
  use them (e.g., the GPU must be able to update textures). \item
  Non-hypervisor memory is not integrity-protected when off-chip due to
  limited on-die memory for code page hashes~\cite{steil_xbox_2008}.
\end{itemize}
Memory encryption is seeded with fresh random values on every
boot~\cite{noauthor_xbox_2022} to prevent replay attacks.

In sum, security was greatly improved on the 360. The major weakness
from the original Xbox, the off-chip 1BL, was eliminated. More money was
invested on custom parts to support the system's security. Strong
ciphers were used. The engineers attempted to ward off DMA
vulnerabilities, which had never even affected the original Xbox.

\textbf{JTAG/SMC: a physical method for exploiting a hypervisor
vulnerability.} It is ironic that many video gamers equate the term
``JTAG'' with ``hardware modchip'' because JTAG was only a convenient
way to deliver a software exploit: ``a new way to exploit the well-known
4532 kernel''~\cite{noauthor_jtag_nodate}. Hardware that Microsoft had
not deemed security-critical could be used to deliver the exploit, but
when the software vulnerability was closed, JTAG/SMC modchips no longer
worked.

The exploit behind JTAG'd consoles was known as the ``King Kong'' attack
because the original delivery mode was to make the King Kong video game
load a malicious saved game file. A memory error in the game could cause
a GPU \ac{dma} operation to modify system data structures in RAM.
Details of the King Kong exploit are out of scope (interested readers
may refer to~\cite{steil_xbox_2008, 15432_protecting_2020_part1}), but
its root causes are relevant. A hypervisor bug, where the bottom 32 bits
of a register were sanitized but all 64 bits were used, made it possible
to coax the hypervisor into using unencrypted, attacker-controlled
memory.

The JTAG/SMC hardware attack was devised because attackers felt that
loading the King Kong game every time they wanted to launch Homebrew was
too \emph{inconvenient}. The hardware exploit worked as follows: the
\ac{smc} firmware was unprotected in NAND flash (a \underline{Class 1}
vulnerability) and could be modified to load components of the King Kong
exploit into registers that defined a \ac{dma} \emph{payload}. However,
the \ac{smc} could not write the register that determined the
\emph{address} where the \ac{dma} payload would be written. The GPU,
which had active JTAG and an exposed JTAG header on the \ac{pcba}, could
be made to write the DMA address register (a \underline{Class 5}
vulnerability, in this context) but could not be made to \emph{initiate
the compromising \ac{dma} transaction} because the CPU disabled the
GPU's JTAG (thereby taking away attackers' control) early in boot,
before the kernel was loaded into main memory. Thus, the \ac{smc} loaded
the payload data, the GPU was instructed to load the DMA address
register using its JTAG interface, and the \ac{smc} initiated the
malicious \ac{dma} at the proper time~\cite{noauthor_jtag_nodate}.

In any implementation of the JTAG/SMC attack, it was necessary to modify
the contents of NAND flash to patch the SMC firmware. The hardware
modification that manipulated the GPU's JTAG signals could be
implemented either as (a)~a microcontroller modchip, or (b)~a few
resistors/diodes connecting Southbridge \ac{gpio} pins to the GPU's JTAG
header. The latter implementation cleverly used free Southbridge pins to
minimize additional hardware required for the attack.

\textbf{Exploiting timing side channels to guess an HMAC digest.}
Another physical attack on the Xbox 360, the ``timing attack'', bypassed
fuse checks to downgrade consoles to a King-Kong-vulnerable hypervisor.
This could be accomplished by modifying the \ac{ldv} associated with the
base kernel~\cite{15432_protecting_2020_part1,noauthor_jtag_nodate,
noauthor_xbox_2007} (see Figure~\ref{fig:xbox360-boot-process}) so that the
4BL would believe the base kernel was up-to-date and launch it. However,
changing the base kernel's \ac{ldv} caused an HMAC check to
fail~\cite{noauthor_xbox_2007}. This check was performed against a value
stored in the \ac{2bl}'s ``pairing data'' header
field~\cite{free60_bootloaders_nodate, noauthor_jtag_nodate} which was
not protected by a signature\footnote{It could not be protected by an
asymmetric signature like the rest of the \ac{2bl} because the pairing
data was console-unique.} and could be modified. However, attackers
could not compute the correct hash for the modified \ac{ldv} value and
update the pairing data because the hash incorporated the secret
console-unique CPU key.

Unfortunately, the hash check was performed by a \texttt{memcmp}
function that worked byte-by-byte, returning an error as soon as a
difference was found. So, by changing one byte of the pairing data hash
and measuring the time it took for \texttt{memcmp} to fail,\footnote{The
time difference between a valid and invalid byte is about 2200
microseconds~\cite{noauthor_xbox_2007}.} it was possible to guess the
correct hash value byte-by-byte. This implementation error amounts to a
failure to correctly verify data stored off-chip: a \underline{Class 1}
vulnerability. A microcontroller could be attached to the \ac{pcba} to
repeatedly 1) reflash NAND with a new base kernel and header that
contained a hash guess, 2) reset the CPU, 3) measure how long it took
for hash comparison to fail, and 4) repeat the procedure with a new
guess. This procedure could bypass fuse checks within a
day~\cite{15432_protecting_2020_part1,copetti-xbox360,
noauthor_jtag_nodate}. Once the hash was found, attackers could launch a
vulnerable kernel and run software to learn their CPU key, enabling them
to arbitrarily modify the 4BL and later boot
stages~\cite{15432_protecting_2020_part1,copetti-xbox360,
noauthor_jtag_nodate}. Codes output on the console's \ac{post} out pins,
which toggled at known points in its boot process, were used as timing
references~\cite{noauthor_jtag_nodate}.

\textbf{Glitch attacks.} The third type of compromise affecting the 360
was fault injection. These so-called \ac{rgh} attacks used FPGA modchips
to inject faults that skipped over buffer comparison failures in the
360's \ac{2bl}. At the moment when hashes were being compared to
determine whether software was authentic and should be booted, the
modchip pulsed the reset signal. This caused the chip's reset procedure
to begin, but the pulse was so short that the procedure did not complete
and the CPU continued executing~\cite{15432_protecting_2020_part3}.
However, the glitch caused the memory comparison result to read as zero,
indicating that the comparison had found no differences even though the
4BL was tampered. Boot proceeded into attacker code. The \ac{rgh}
attacks are \underline{Class 1}: they exploit implementation errors that
undermine secure boot.

Different console motherboard revisions required different variants of
the \ac{rgh}. In all variants, the first step was to slow the 360's CPU
because the reset pulse had to be sent at \emph{exactly} the right
moment and the Xbox's native speed was too fast for modchips to glitch
accurately. For early console revisions, this was accomplished by
asserting the CPU's \texttt{PLL\_BYPASS}
signal~\cite{15432_protecting_2020_part3, noauthor_xbox_2022}, which was
exposed on the \ac{pcb}, causing the CPU to execute at the frequency of
its external oscillator. This bypass signal was removed in later console
revisions, but hackers discovered that the HANA peripheral, which
controlled the CPU clock frequency, would reduce the clock speed in
response to I2C commands~\cite{15432_protecting_2020_part3,
noauthor_xbox_2022}.

Next, the reset pulse was sent. The reference point for determining when
to send the pulse was, as in the HMAC timing attack, the CPU's
\ac{post} output pins~\cite{15432_protecting_2020_part3}.

The glitch attack was not perfectly stable and, if it failed five
attempts in a row, the console would stop boot and display an error
message. However, this behavior could be bypassed by modifying the SMC
code in a similar manner to the JTAG/SMC
hack~\cite{15432_protecting_2020_part3}.

Microsoft soon added random delays and redundant checks in their
\ac{2bl} that closed the glitch vulnerability, but weaknesses connected
to manufacturing boot modes and ciphertext malleability allowed
adversaries to revert their console's firmware to \ac{rgh}-vulnerable
versions~\cite{15432_protecting_2020_part3}. Microsoft tried removing
the \ac{post} debug output traces on later motherboard revisions in an
attempt to stymie adversaries by removing their timing reference, but
the hackers discovered that the \ac{post} pins on the CPU were still active
and were on the outermost ball ring of the BGA CPU package, so they
could still be accessed~\cite{15432_protecting_2020_part3}. A plastic
part with springloaded pins that fit around the CPU and contacted the
\ac{post} pins was sold to make this process easier. The HANA chip that
attackers were using to reduce the CPU frequency was removed from the
motherboard and integrated in the South Bridge, but the adversaries
learned how to talk to that same logic in its new home and added an
oscillator to their modchips to keep \ac{rgh}
working~\cite{15432_protecting_2020_part3}. 

In the final Xbox 360 motherboard revision, Microsoft finally killed
\ac{rgh}. Some speculate that they added or improved logic on the CPU
die to filter the reset line. They also disabled \ac{post} signals
(instead of merely removing the traces that connected to them) to remove
the attackers' timing reference. This final 360 hardware revision is
widely considered secure against physical attacks. Interested readers
may refer to~\cite{15432_protecting_2020_part3} for more details on the
cat-and-mouse match of \ac{rgh} attacks.

\textbf{An insufficiently-secure security-critical peripheral.} In
contrast to all of the attacks described above, which targeted the
Xbox's CPU, the final class of Xbox 360 \ac{pcba} attacks exploited its
disc reader. The Xbox protected itself against code from discs in
several ways: game data is signed and can only be executed after a
signature check, and the hypervisor limits the damage from
vulnerabilities in authentic games. However, integrity of the
\emph{disc} (in contrast to the \emph{data} on the disc) has to be
delegated to the disc reader; the CPU has no way to tell whether the
piece of plastic inside the disc reader is authentic. Thus, attacks on
the disc drive could not run arbitrary software (e.g., Linux) but did
enable pirated games.

Xbox 360 games were commodity dual-layer discs upgraded with extra
security sectors on inner rings~\cite{15432_protecting_2020_part1,
noauthor_xbox_nodate} which were out of range for standard disc drives.
Several companies were contracted to make drives for the Xbox 360. All
implemented the same interface so that they were interoperable. For our
discussion, two features of this interface are important. First, the CPU
could issue programming/erasure commands to the disc drive controller.
Second, when the CPU fetched anti-piracy data from the disc drive, the
exchange was encrypted with a console-unique key. Other command and data
exchanges were not encrypted, so it was possible for attackers to send
commands to the drive.

Attempts to secure communications with the disc reader were undermined by
\underline{Class 1} and \underline{Class 2} vulnerabilities in the disc
drive controller.

The disc drives on early Xbox 360 generations were trivially broken. One
early disc reader stored its drive key unencrypted in a discrete flash
chip on the drive
\ac{pcba}~\cite{15432_protecting_2020_part1,copetti-xbox360}, a
\underline{Class 2} vulnerability. The drive controller also did not use
cryptographic firmware
verification~\cite{15432_protecting_2020_part1,copetti-xbox360}, a
\underline{Class 1} vulnerability. Attackers dumped the key, programmed
a drive controller with a firmware that enabled piracy, and re-paired
the modified drive firmware to their CPU using the dumped key.

Another drive type released in the first Xbox generations was slightly
more secure; its firmware was encrypted and integrity-checked. However,
attackers quickly discovered that the drive had a manufacturing
mode~\cite{15432_protecting_2020_part1} which could be entered by
presenting a special boot disk or shorting a few pins together. It was
possible to overwrite the drive's firmware from the manufacturing mode
so that it would play copied games just like the other family of drives.
The manufacturing mode opens another \underline{Class 1} vulnerability.

A new drive revision was released that, for a time, seemed more secure:
its flash memory die was in the same package as the drive controller so
that the firmware could not be dumped or examined, and there was no
known way to make this revision enter a service mode. However, attackers
soon discovered that if they opened the drive half-way while power was
disconnected and then reconnected power, \emph{the drive would spit out
its key over UART}. Initially, adversaries exploited this weakness by
buying first-generation drives (which were easy to re-flash) and
programming them with 1) the dumped key and 2) the logical name from the
newer drive generation so that the CPU would not panic at the drive type
change. However, all of this trouble with drive replacement went away
when attackers figured out they could get the new drives to enter
maintenance mode (from which they would accept a bogus firmware update)
by issuing commands to the drive controller to make it erase its
internal flash~\cite{15432_protecting_2020_part2}.

A drive revision was quickly released that would not dump its key over
UART; it dumped its key over SATA
instead~\cite{15432_protecting_2020_part2}. Allowing secrets outside an
\ac{ic} trust boundary is a \underline{Class 2} vulnerability.

Finally, a drive revision was released that did not have a key dump
feature. Adversaries correctly suspected that the disc reader would
still enter maintenance mode if it believed its internal flash contained
no firmware. To trick the drive into entering maintenance mode without
erasure, attackers shorted power to ground via a resistor so that the
voltage supplied to the drive controller was so low that it could not
read its internal flash properly. The controller could see only
\texttt{0xFF} bytes when it attempted to read from flash, so it assumed
flash was empty and dropped into maintenance mode, where hackers could
instruct it to dump its internal
flash~\cite{15432_protecting_2020_part2}. Eventually, another strange
power sequence was discovered that would enter maintenance mode by
grounding pins on the board to make drive key dumping
easier~\cite{15432_protecting_2020_part2}. Relying on attackers'
inability to enter manufacturing mode instead of cryptographically
verifying firmware opens a \underline{Class 1} vulnerability.

The next drive revision had no security whatsoever, just like the days
when the console was first released. It was hacked
immediately~\cite{15432_protecting_2020_part2}.

In a later drive revision, Microsoft attempted to thwart drive
reprogramming by bonding the flash die's Write Protect pin to ground
inside the controller's package, but they still did not
cryptographically verify firmware. This bond wire was positioned such
that a careful adversary could cut it \emph{by drilling into the drive
controller package}. Then, they could use the old drive-half-open trick
to drop into manufacturing mode and reprogram the flash as
usual~\cite{15432_protecting_2020_part2, chen_guarding_2019}. This
\underline{Class 1} exploit was aptly called the ``Kamikaze
attack''~\cite{ashcraft_one_2021}. Special kits that assisted
inexperienced modders with the positioning and depth of their drill were
commercialized~\cite{chen_guarding_2019}.

The next drive revision was not susceptible to firmware dumping or
overwriting tricks and would not spit out its drive key. However,
adversaries who managed to hack their CPUs (e.g., using \ac{rgh}) could
still extract their drive key and either (a) replace their drive
controller \ac{pcba} with an exact copy that had an
unlocked\footnote{That is, a version of the chip that had not yet been
programmed with a vendor's firmware signing public key.} version of the
controller \ac{ic} or (b) purchase an unlocked \ac{ic} and replace the
chip on their original drive \ac{pcba}. In either case, a new firmware
patched with the drive key from the CPU was flashed to the unlocked
controller to defeat disc reader
security~\cite{15432_protecting_2020_part2}. \Ac{rgh} remains a
\underline{Class 1} vulnerability of the CPU, as dumping the drive key
required hijacking the CPU.

The cat and mouse game of disc drive attacks finally ended when
Microsoft shipped consoles where neither the CPU nor disc drive would
cough up the drive key~\cite{15432_protecting_2020_part2}: the final
drive revision, which would not drop into manufacturing mode to allow
firmware replacement or dump its keys, was paired with a CPU that was
not susceptible to \ac{rgh}.

\subsubsection{There weren't any Xbox One modchips} In Microsoft's
latest consoles, strong digital security architecture and careful
implementation appear to have succeeded against \ac{pcba} tampering
adversaries~\cite{chen_guarding_2019}.

\textbf{Distrust everything outside a trusted IC.} A \ac{soc} is the
console's root of trust and signals from every other component on the
board are treated with suspicion~\cite{chen_guarding_2019}. The only
other security-critical \ac{ic}, the disc reader, uses proper secure
boot~\cite{chen_guarding_2019} and incorporates all of the
implementation lessons that Microsoft and its vendors learned from disc
drive attacks on the Xbox 360.

\textbf{Dedicated security coprocessor.} One of the security IPs in the
Xbox One's \ac{soc} is a security coprocessor that operates on all of
the console's secrets and does not permit any other IP in the \ac{soc}
to read them under any circumstances~\cite{chen_guarding_2019}. This
corrects a major weakness in the Xbox 360: the CPU could read important
secrets, enabling any adversary who managed to hijack the CPU to learn
them. Besides isolating secrets from the CPU, the security core has
dedicated communication channels with the CPU and RAM encryption engine
that prevent other IP blocks from interacting with the security engine
in unintended ways (i.e., communications with the core would never
appear on common buses).

\textbf{Hypervisor with stronger VM isolation.} Like the 360, the Xbox
One uses a hypervisor to check code authenticity and integrity before
marking any page executable. However, the Xbox One hypervisor runs
games, system functions, and hardware drivers in separate virtual
machines~\cite{chen_guarding_2019} whose memory is encrypted using
different keys~\cite{martin_amds_2019,
chen_guarding_2019}.\footnote{This architecture is now deployed in
server-class CPUs to strengthen isolation between tenant
\acp{vm}~\cite{amd_amd_nodate, martin_amds_2019}.} To achieve this, the
security coprocessor loads a different key into the RAM encryption
engine for each \ac{vm} in the system~\cite{amd_amd_nodate}. Per-\ac{vm}
encryption prevents software vulnerabilities in one \ac{vm} from
affecting any of the others: any values written to DRAM under one
\ac{vm}'s key decrypt as garbage when read by a different \ac{vm}. In
particular, this design prevents game or OS vulnerabilities from coaxing
the hypervisor into using unencrypted memory as the Xbox 360 King Kong
attack did.

\subsection{The NSA ANT Catalog} \label{sec:nsa}

Hardware modification of IT equipment in transit (``interdiction'') is
another common motivation for recent work. Examples of citing papers
include, but are not limited to:
\cite{10.1145/3401980,boone_tpm_2018,bhattacharyay_automated_2022,10.1145/3588032,guo_mpa_2017,zhu_pcbench_2021}.

The NSA ANT Catalog is a leaked U.S. government document that describes
several hardware implants available to U.S.
intelligence~\cite{appelbaum_nsas_2013}. According
to~\cite{glenn_greenwald_how_2014}, some were installed by interdiction.
The NSA
Playset~\cite{dominic_spill_nsa_2015,fitzpatrick_tools_2014,
michael_ossmann_nsa_2022,joe_fitzpatrick_tao_2016} is an effort by
hardware hackers to make open-source versions of the ANT Catalog
attacks. Others have also studied the technical feasibility of these
implants~\cite{wakabayashi_feasibility_2018,salihun_nsa_2014_pt1,
salihun_nsa_2014_pt2,stanislav_mechanics_2014}.

Even the nation-state spy implants described in the ANT Catalog can,
with a few exceptions, be mitigated by security architecture or
implementation fixes. As the NSA Playset's authors put it, these
implants are the result of ``design flaws'' exploitable by ``12-year
olds''~\cite{fitzpatrick_tools_2014}.

\textbf{Abuse of active debug infrastructure.} One NSA implant attaches
to a server motherboard's debug infrastructure to install malware on the
CPU. Details of the malware are irrelevant; active debug interfaces are
widely understood to be a vulnerability if physical attacks are in
scope~\cite{stanislav_mechanics_2014, rosenfeld_attacks_2010}. The NSA
Playset has demonstrated that a wide variety of software payloads can be
installed using the same JTAG Trojan: they have changed file permissions
in Linux~\cite{fitzpatrick_nsa_2015}, bypassed Linux login password
checks~\cite{joe_fitzpatrick_tao_2016}, and manipulated outputs of an
industrial PLC~\cite{joe_fitzpatrick_tao_2016}.

Most microprocessors are capable of disabling JTAG infrastructure
permanently via eFuse or write-protected flash configuration bits. JTAG
is helpful for firmware development, but it should always be disabled in
deployed systems to prevent easy, catastrophic attacks. Failure to do so
could lead to many classes of vulnerabilities, depending on precisely
how the active debug infrastructure is abused.

\textbf{Protection against malicious peripherals.} The Catalog describes
several implants installed in USB cables or host ports that inject
software exploits and include a radio transceiver to bridge air gaps.
System protections cannot prevent transceivers being installed to bridge
airgaps, but software and USB driver vulnerabilities are garden-variety
\underline{Class 5} implementation errors~\cite{hernandez_firmusb_2017}.

Another implant replaces hard drive firmware. The Catalog gives the
impression that the implant exploited vulnerabilities in master boot
record (MBR) parsing. If this is true, the underlying issue is the
software vulnerability, not the hardware delivery mode. However, the
more important physical vulnerability is an untrustworthy peripheral: if
a system component is unauthenticated, all bets are off.
\cite{zaddach_implementation_2013} illustrates how a hard disk without
protected boot can modify written data or create a backdoor for
exfiltrating information. Firmware signing is listed in the paper as an
effective countermeasure~\cite{zaddach_implementation_2013}. As the
authors point out, firmware signing does not prevent complete
replacement of the hard drive, but authenticating the peripheral \ac{ic}
closes this gap. Yet another attack leverages a malicious hard drive
firmware to turn the hard drive into a microphone: the malicious
firmware keeps track of ``position error measurements'' of the disc read
head, which is highly sensitive to vibrations such as those caused by
nearby speech, and using them to reconstruct an audio
signal~\cite{kwong_hard_2019}. This is cool! But we reiterate: if
\ac{pcb} attacks are in-scope, system architects should authenticate
peripheral \acp{ic} to mitigate \underline{Class 5} vulnerabilities and
should be very careful incorporating any peripherals that do not check
signatures on their firmware, i.e., that have \underline{Class 1}
vulnerabilities.

\textbf{Snooping and spoofing ethernet traffic.} One NSA implant lives
in ethernet jacks where it snoops traffic and injects packets onto a
target network. It is coupled with an RF transceiver for remote control
and crossing airgaps. The NCC Group~\cite{davis_dock_2013} prototyped a
similar attack.

An implant in an ethernet jack `taps the wire' in the most traditional
sense: a processor on the \ac{pcba} is the cryptographer's ``Alice'', a
remote computer is ``Bob'', and the implant is  ``Eve'' if it is passive
or ``Mallory'' if it injects packets. These implants are precisely the
adversaries that cryptographic protocols were designed to protect
against. Any connection using industry-standard security (e.g., TLS)
closes the \underline{Class 2} vulnerabilities associated with lost
secrets, the \underline{Class 3} vulnerabilities associated with failing
to authenticate a remote computer, and the \underline{Class 4}
vulnerabilities from not integrity-protecting communications.

\textbf{DMA attacks by PCI(e) devices.} A fifth NSA implant was
installed ``plug-and-play'' style into a PCI adapter to bridge airgaps
and install malware on the host. The NSA Playset prototyped this
capability with an off-the-shelf PCI development
kit~\cite{fitzpatrick_tools_2014} and demonstrated a password check
bypass on then-modern Macs. The password bypass modifies code in RAM so
that the password verification function accepts any non-empty password
as correct~\cite{inception}. Later, PCILEECH~\cite{frisk_direct_2016}
achieved similar payloads with higher read and write speeds.
\cite{frisk_direct_2016} is still maintained as a penetration testing
tool.

As explained in~\cite{ruxcon-hit-by-bus,heasman_implementing_2007},
older computers typically imposed no limits on memory accessible to DMA
peripherals, so DMA devices could read or modify critical code or data
in RAM. The Windows 9x, 2000, 20003, XP, or Vista systems that this NSA
implant targeted almost certainly fell into this category. These attacks
are partially mitigated by enabling an
\ac{iommu}~\cite{frisk_direct_2016}, a hardware block built into most
modern processors~\cite{virtualization-security,apple-iommu} that
restricts which memory is available to different I/O devices.
\acp{iommu} prevent \underline{Class 5} ``plug-and-play'' attacks, but,
similar to other blacklisting-based mitigations, they can be bypassed by
\ac{pcba} attacks located after the \ac{iommu}. The real vulnerability
at the root of DMA attacks are in \underline{Class 4}: sensitive data
written to working memory by the CPU are neither encrypted nor
integrity-checked. Though we do not know of any examples, it is also
easy to imagine \underline{Class 2} vulnerabilities from writing
unprotected secrets to RAM that get Hoover'd up by a DMA implant.

\textbf{Retroreflectors.} The Catalog includes several
``retroreflectors'', which use a signal of interest to modulate radar
waves broadcasted by an attacker who monitors the reflected waves from a
distance. The modulations in the reflected waves allow recovery of the
signal of interest without high-power
transmitters~\cite{wakabayashi_feasibility_2018,michael_ossmann_nsa_2022}.
ANT Catalog implants can broadcast digital data, analog video,
ambient audio, or simply act as a location beacon. Green Bay
Professional Packet Radio has a series of YouTube videos that analyzes
and demonstrates their principles of operation~\cite{gbppr}.

The NSA Playset has prototyped retroreflectors that work with commercial
software-defined radios. Their PoCs can monitor PS/2 keyboards, low- to
mid-speed USB, and, to a limited extent,
VGA~\cite{michael_ossmann_nsa_2022}. A few years later,
\cite{wakabayashi_feasibility_2018} took a second look at the attack and
affirmed retroreflector viability at short range with a PoC attack on a
USB keyboard. Both used a rudimentary setup and simple signal recovery;
more sophisticated analysis could increase range and accuracy.

Systems defenses cannot prevent devices that broadcast ambient audio or
act as a beacon; these devices leech power from a host system but do not
otherwise interact with it. These are our first \underline{Class 6}
attacks. The retroreflectors that broadcast digital data are
\underline{Class 2} problems that can be mitigated by encrypting signals
between digital components.\footnote{These attacks are \underline{Class
3} \emph{as proposed by the NSA Playset and Catalog}, but keyboards,
mice, and other digital or analog sensors generally create
\underline{Class 6} vulnerabilities. For example, encryption between a
keyboard and host can rule out trojans in USB cables, but the keys
themselves are simple pushbuttons that cannot be protected by system
security.} The retroreflector that broadcasts analog video is more
difficult to address because there are no accepted solutions for analog
signal encryption; this vulnerability is \underline{Class
6}.\footnote{Note that modern digital video signals are typically
encrypted, e.g., using HDCP on HDMI~\cite{kuhn_electromagnetic_2005}. A
similar attack on \emph{digital} video would fall into \underline{Class
2}.}

\subsection{Attacks on TPMs} \label{sec:attacks-on-tpms}

\Acp{tpm} are dedicated chips\footnote{Some \acp{soc} incorporate
\ac{tpm} logic on-chip. There are also ``soft'' TPMs. But our discussion
is only concerned with discrete \acp{tpm}.} for storing secrets, making
digital signatures, generating entropy, and other security-critical
operations. Sometimes \acp{tpm} are used just to isolate these
operations from potentially vulnerable software on a CPU. However,
another usage model, `remote attestation', relies on \acp{tpm} as
trustworthy reporters of system integrity in untrusted systems.

In the attestation model, the first code to run on a CPU gives the
\ac{tpm} a hash (or ``measurement'') of the second firmware stage into
the \ac{tpm}'s \acp{pcr} before launching it. Similarly, each subsequent
boot loader writes the \acp{pcr} with a hash of the next stage so that
when boot is finished, the \ac{tpm} holds a record of all loaded
software. The \ac{tpm} ensures that the \acp{pcr} cannot be modified or
cleared without a reset so that malware cannot tamper with attestation
records. Later, when a remote party wants to know whether to trust a
device (e.g., trust it to decrypt DRM-protected media without allowing
pirates to copy the plaintext), they can ask the device for its \ac{tpm}
hashes. The \ac{tpm} signs the contents of its \acp{pcr} with an
``endorsement'' key~\cite{tcg_tpm_2011} that proves the measurements are
held in an authentic \ac{tpm} (and are therefore trustworthy), then the
measurements and signature are returned to the verifier.

The \ac{tpm} threat model centers on keeping secrets: endorsement keys
must be protected so that attestation is meaningful, and keys entrusted
to the \ac{tpm} by a system must also be protected. Additionally, it
must not be possible to make a \ac{tpm} tell a remote verifier that a
system is running `safe' software when it is not, and, if a processor
relies on a \ac{tpm} for strong entropy, received entropy must not be
compromised.

Researchers have prototyped several \ac{pcba} attacks against TPMs: 1)
an attack on \ac{tpm}-supplied entropy, 2) a snooping attack that
recovers cryptographic keys stored in \acp{tpm}, and 3) two attacks on
remote attestation.

\textbf{Implant-supplied randomness.} In platforms that rely on a
\ac{tpm} to generate cryptographically secure random numbers, a hardware
interposer could respond to CPU requests for random numbers with known
values if CPU$\leftrightarrow$\ac{tpm} communications are not
integrity-protected, thereby undermining cryptographic protocols that
require strong entropy~\cite{boone_tpm_2018}. If \ac{pcba} security is
part of a system's threat model, this is a \underline{Class 4} security
architecture failure.

\textbf{Snooping cleartext secrets off a bus.} In the default
configuration, most \acp{tpm} do not encrypt traffic between the
\ac{tpm} and CPU, allowing adversaries with inexpensive equipment to
sniff keys off the bus. \cite{denis_andzakovic_extracting_2019}
describes how this attack could retrieve disk encryption keys.

If \acp{tpm} are only intended to protect secrets from CPU software
vulnerabilities, transmitting keys in the clear between the CPU and
\ac{tpm} is acceptable. But if the \ac{tpm} is supposed to, e.g.,
protect disk encryption keys in the event that a laptop is stolen, then
transmitting secrets in cleartext across the \ac{pcba} is a
\underline{Class 2} security architecture gap.\footnote{\ac{tpm} v2
standards offer optional safeguards against PCBA tampering by encrypting
and authenticating the channel between a CPU and
\ac{tpm}~\cite{boone_tpm_2018}, but these features are optional and,
according to \cite{boone_tpm_2018}, were almost universally absent from
TPM drivers as of 2018.}

\textbf{Assumptions that do not hold for \ac{pcb} adveraries.} Remote
attestation relies on a handful of assumptions, one being that \ac{tpm}
initialization is always performed by a piece of trusted code that runs
immediately after a CPU is reset~\cite{tcg_tpm_2011}. In other words, it
is assumed that the \ac{tpm} and CPU will always be reset together. A
physical adversary can violate this assumption by electrically isolating
the \ac{tpm}'s reset signal from the CPU's. This allows the attacker to
reset the \ac{tpm} and re-initialize it with forged measurements to
deceive remote verifiers~\cite{kauer_oslo_2007}. This is known as a
``TPM reset attack''.

\cite{winter_hijackers_2013} discusses how, with the help of a \ac{pcba}
implant, \ac{tpm} reset attacks can also be applied in the reverse
direction: if a CPU is reset but its \ac{tpm} is not and the CPU's early
boot stages are prevented from communicating with the \ac{tpm}, the
\ac{tpm} will still contain software measurements from the previous
boot,  enabling untrusted software to deceive a remote verifier into
trusting a device based on stale measurements. The \ac{pcba} implant
prevents communication with the \ac{tpm} by unconditionally asserting a
signal on the CPU $\leftrightarrow$ TPM bus.

At first glance, reset attacks look like they cannot be countered by
digital architecture: system security primitives cannot prevent two
discrete chips from being reset separately. However, closer examination
reveals that separate resets are a distraction rather than the main
challenge. The real problem is in \underline{Class 3}: \acp{tpm} assume
they are initialized by a trusted device but they do not authenticate
that device. A more robust solution could leverage \ac{pki} to enable
\acp{tpm} to verify they are being initialized by an authorized device,
e.g., a genuine CPU from a trusted manufacturer.

\Ac{drtm}, which is discussed next, was once thought to be a solution to
reset attacks~\cite{kauer_oslo_2007}. However, the remainder of this
section explains that \ac{drtm} is also vulnerable to physical
adversaries.

\textbf{Lack of integrity protection enabling packet-in-packet attacks
on D-RTM.} There are two flavors of \ac{tpm}-based attestation. So far,
our discussion has focused on \emph{static} \ac{tpm} attestation. Static
attestation has the drawback that verifiers must maintain lists of
attestation values associated with software and hardware configurations.
Such a list quickly becomes unwieldy because a typical PC loads many
pieces of device-specific software during boot and there may be multiple
authentic versions of each boot loader stage (e.g., due to updates).
\textit{Dynamic} attestation,\footnote{It is called ``dynamic''
measurement because it occurs on a running system without requiring a
system reset~\cite{tcg-drtm}.} or \ac{drtm}, aims to solve this problem
by allowing untrusted devices to transition to a trusted state via a
short chain of hashes that is common across many platforms.

A \ac{drtm} launch begins when control of the CPU is transferred to
trusted code via a special CPU instruction. This code sets up an
environment that is protected from untrusted program snooping or
interference~\cite{tcg-drtm}. Sensitive data is then processed only
within this secure environment.

The \ac{drtm} launch sequence gets recorded with a special set of
\acp{pcr}. Restrictions on writing \ac{drtm} \acp{pcr} are
hardware-enforced: on Intel chipsets, for example, the South Bridge
drops any writes to the \ac{drtm} \acp{pcr} that do not originate from
the \ac{drtm} launch code~\cite{winter_hijackers_2012}. \Ac{drtm}'s
security rests on the inability of `normal' software, including
privileged software running on the CPU, to write these special
\acp{pcr}~\cite{tcg-drtm, winter_hijackers_2013, winter_hijackers_2012}.

Unfortunately, \cite{winter_hijackers_2013}~demonstrates that a hardware
implant can transform a seemingly innocuous bus transaction into a
\ac{drtm} initialization sequence after hardware blacklisting has
already passed. The authors demonstrate attacks against both LPC bus
\acp{tpm} (using the \texttt{LFRAME} signal) and I2C bus \acp{tpm} (by
manipulating the I2C clock). These attacks mount a packet-in-packet
attack: they build an LPC or I2C transaction where the first half is a
trigger sequence that the hardware implant will recognize as a cue to
break the malicious packet by manipulating bus signals, and the second
half is a \ac{drtm} register extension command.

These \ac{drtm} attacks are enabled by a straightforward
\underline{Class 4} vulnerability: communications between the CPU and
\ac{tpm} must be integrity-checked to prevent a man-in-the-middle from
mangling messages. Also, as with the Xbox jam tables, we see that
blacklists are not a good security strategy.

\subsection{More attacks} This section reviews \ac{pcba} implant
research that is not part of a `family' -- these attacks don't target a
common system (e.g., TPMs, game consoles, etc.), and they were not
perpetrated by a single party like the ANT Catalog.

\paragraph{General purpose computers not intended to be secure against
PCBA adversaries.}

\cite{davis_dock_2013} crams a single-board computer with a 3G modem
into a hollow nook in a laptop dock and solders in circuits to snoop or
modify signals from dock peripherals. It can grab or inject keystrokes
on the USB bus (\underline{Class 2}), record video frames from a USB
webcam (\underline{Class 2}) or an analog monitor (\underline{Class 6}),
snoop or inject ethernet traffic (\underline{Class 2/4}), snoop general
USB traffic (\underline{Class 2}), and record
audio~\cite{davis_dock_2013} (\underline{Class 6}). Although there's not
much that can be done for analog audio streams or video signals, the
rest are solvable with systems approaches. The issues of keystrokes and
ethernet taps were discussed in Section~\ref{sec:nsa}. Digital video and
webcams can use standard encryption between \acp{ic} to mitigate
tampering.

Voting machines from the early 2000's are another example of computers
that were not designed to withstand physical attacks.
\cite{feldman_security_2007, gonggrijp_studying_2007} both found that
leading voting machine manufacturers had no boot security in place
whatsoever (\underline{Class 1}). \cite{feldman_security_2007} shows
that, in their machine, the boot address mapped to one of three memory
chips on the board, any of which could be reprogrammed or replaced by an
adversary. For example, \cite{gonggrijp_studying_2007} replaces one
voting machine's firmware with a chess program. DefCon teardowns of
other voting machines have found that similar attacks are
possible~\cite{blaze_report_2017}. Both papers suggest cryptographic
boot verification as a mitigation against tampered flash
chips~\cite{feldman_security_2007, gonggrijp_studying_2007}.

\paragraph{Insecure communication with a critical peripheral.}
\cite{fietkau_swipe_2018} reverse engineers a fingerprint smart card and
shows that, by interposing between the fingerprint sensor and the CPU,
it is possible to replay old fingerprints, ``brute force the matcher''
by injecting many images of fingerprints, and more.
\cite{chen2023bruteprint} does the same on real-world smartphones. This
was all possible because signals between the sensor and CPU were
unencrypted and not integrity protected, \underline{Class 2} and
\underline{Class 4} vulnerabilities. The researchers state that their
\ac{pcba} attacks all boil down to a failure to use standard security
architecture. Note that it is also important to verify the authenticity
of critical peripherals like fingerprint sensors to prevent
\underline{Class 3} vulnerabilities. The attackers in these instances
simply didn't need to get that creative.

\cite{shwartz_shattered_2017} demonstrates two attacks that can be
launched by a ``Trojan'' smartphone screen: 1) the screen exploits
vulnerable device drivers to hijack control flow of the main CPU, and 2)
it logs a user's touches and sends false touch events to the CPU. The
screen is simply an I2C peripheral: its hardware provenance and
integrity should be established with \ac{pki} to prevent
\underline{Class 3} vulnerabilities, and its firmware should be
authenticated to prevent \underline{Class 1}.\footnote{Ironically,
smartphone makers have tried to authenticate their components using
\ac{pki} but right-to-repair legislation prevents them from securing
consumers against malicious replacement parts.}

\paragraph{Voltage glitching in embedded systems.}
Chip.fail~\cite{thomas_roth_chipfail_2019} is an FPGA development board
connected to a 3-channel switch that enables experimenting with voltage
glitching. The device has successfully attacked several common IoT
microcontrollers, including some with brownout reset (BOR), which was
advertised as a glitching countermeasure, and showed that naive
comparisons in firmware could be bypassed with voltage glitching. They
also achieved flash option byte downgrade on an STMF2 microcontroller,
which re-enables the chip's debug infrastructure. Section~\ref{sec:nsa}
has already discussed the consequences of active debug infrastructure
in deployed systems.

Insofar as glitching is an implementation error that undermines stored
program verification, it is a \underline{Class 1} vulnerability. As
discussed in Section~\ref{sec:xbox-360}, random delays and removal of
timing reference signals make it near impossible for attackers to create
stable attacks. A specific mitigation for the flash option byte
downgrade in STM microcontrollers is to check status registers shortly
after boot and ensure that they read as expected before reading any
sensitive values from RAM~\cite{obermaier_shedding_2017}.

\paragraph{Inducing crosstalk by moving \ac{pcb} traces.}
\cite{ghosh_how_2015} proposes that the location or dimensions of
\ac{pcb} traces could be modified to enable signal changes on one trace
to impact a neighbor. A simulation is conducted to demonstrate that
doubling a trace's width and introducing a long parallel section in two
traces results in a few volts of crosstalk. The authors do not present a
specific attack PoC, so we reason about how the attack might interact
with a full system. If the victim trace is responsible for data
transmission, this is a \underline{Class 4} vulnerability that can be
mitigated by checksums. If the victim is a different kind of trace~--~an
input that triggers a processor interrupt (\underline{Class 6}), a
discrete logic input or output (\underline{Class 6}), or an analog
control signal (\underline{Class 6}), then systems security has no ready
answer.

\cite{cottais_second_2018} proposes that various
hardware modifications could be used to influence a radio transmitter to
add a covert ``polyglot'' transmission to a legitimate radio signal. The
modifications influence the RF frontend's oscillator frequency using crosstalk
to add an extra modulation on top of a legitimate signal. The legitimate
signal can still be demodulated. Digital security primitives cannot
prevent modulations to an oscillator frequency~--~this vulnerability is
\underline{Class 6}.

\cite{loveless_pcspoof_2023} describes a more specific attack involving
crosstalk that enables a Trojan low-priority device in a \ac{tte}
network to disrupt the network's high-priority traffic. The resulting
loss of synchronization, the authors say, is sufficient to cause
unrecoverable errors in spacecraft, aircraft, automobiles, industrial
control systems, and other real-time systems.

\Ac{tte} enables mixed-criticality communication, meaning it sends
low-priority and high-priority messages on the same physical
infrastructure while guaranteeing that high-priority messages cannot be
impacted by low-priority ones. In particular, special ``\acp{pcf}'' may
only be sent by designated high-priority devices because these are
essential for maintaining synchronization in the network. If a
low-priority device tries to send a \ac{pcf}, the \ac{tte} switches drop
the packet~\cite{loveless_pcspoof_2023}.

However, \cite{loveless_pcspoof_2023} found that it was possible to
modify low-priority traffic when it was on an outbound port
(\emph{after} it had already been relayed by the \ac{tte} switch). This
is accomplished by conducting a high voltage pulse (on the order of
kilovolts) over a low-priority ethernet cable into a port on the
\ac{tte} switch. The electrical design of ethernet ports, which includes
galvanic isolation, ensures that the victim port will not be destroyed
by the high voltage pulse, but the pulse will cause significant EMI that
impacts adjacent switch ports. If an apparently-benign low-priority
packet is being transmitted out of the switch at the same instant that a
high voltage pulse induces EMI on that outbound port, a connection reset
event may occur on the line, enabling a packet-in-packet attack. The
`inner packet' could be a valid \ac{pcf}, and the authors demonstrate
that a \ac{pcf} generated in this manner has a good chance at causing
unsafe desynchronization in realistic \ac{tte}
environments~\cite{loveless_pcspoof_2023}.

The authors suggest various solutions to this vulnerability. The
fundamental flaws are \underline{Class 2/4}, and link-layer encryption
for high-priority traffic is among the authors' suggested
mitigations~\cite{loveless_pcspoof_2023}. Note that this attack is
another case study where blacklisting unwanted behavior fails as a
security strategy.

\paragraph{Tampering analog control circuitry.} \cite{ghosh_how_2015}
adds a resistor, capacitor, and PMOS to an op-amp circuit. These
modifications have the effect of grounding a microcontroller input after
the circuit has been active for a while. This was done in a fan
controller circuit and the grounded pin deactivated the fan. Systems
security has little to say here. This is a \underline{Class 6}
vulnerability.

\subsection{Destructive attacks}

\paragraph{Destructive high voltage.} It is possible to design a Trojan
component that deliberately destroys components on a \ac{pcba}. For
example, \cite{noauthor_usbkill_nodate} is a USB device that applies
large positive and negative voltage to a host's USB ports. It is
marketed as a tool for pentesters. A few hardware design techniques can
provide resilience against extreme voltages, but systems security has
little to offer. This is \underline{Class 6}.

\paragraph{Trace breakage due to electromigration.} \cite{ghosh_how_2015,
mcguire_pcb_2019} discuss attacks where a malicious fabricator thins
\ac{pcb} traces. \cite{ghosh_how_2015} suggests that this could cause
systems to overheat by increasing trace resistance.
\cite{mcguire_pcb_2019} proposes such an attack could induce
electromigration at a (somewhat) predictable point in the future.
Security architecture cannot prevent a too-thin trace from eventually
separating from high current density: this vulnerability is 
\underline{Class 6}.

\section{Conclusions and future directions} \label{sec:conclusion}

We explained how security architecture can mitigate \ac{pcba} tampering
and why its assumptions are reasonable. Then, we considered how it
complements, and overlaps with, tamper detection. Finally, we examined
over fifty real-world \ac{pcba} attacks to show that the \ac{pcba}
attacks most commonly-cited as motivation by recent tamper detection
research can be mitigated by basic security architecture or by fixing
implementation errors. Our discussion of security architecture's role in
\ac{pcba} tamper detection fills a gap in prior work on this topic.

\textbf{Most of the attacks we reviewed affected systems that either did
not implement a \ac{pcba} security architecture, or implemented one that
is obviously inadequate if board-level attacks were part of the system's
threat model.} This set includes the BMC attacked in Bloomberg News'
``The Big Hack'', the original Xbox, \acp{tpm}, computers targeted by
NSA ANT attacks, voting machines, cell phone screens, fingerprint
sensors, and others. \textbf{Except for voltage glitching in embedded
bootloaders, the Xbox and Xbox~360 were the only systems we studied that
tried, but failed, to defend against \ac{pcba} attacks with security
architecture.} The Xbox's vulnerabilities came from ruthlessly-exploited
security architecture oversights and from implementation mistakes.
Crucially, \textbf{improved implementation and architecture in later
Xbox~360s and the Xbox One seem to have mitigated tampering}, as there
are no known tampering attacks against these
consoles~\cite{chen_guarding_2019}.

This raises an important question: \emph{why have so few systems used
security architecture to protect themselves from \ac{pcba} attacks?} We
have no special insight into decisions made by most of the companies
whose systems were compromised in our survey, but the \ac{tcg} tells us
clearly that security against \ac{pcba} tampering was not an objective
of the v1 \ac{tpm} protocol~\cite{lawson_tpm_2007,tcg_tpm_2011}. In
light of this, it is no surprise that \ac{pcba} attacks on \acp{tpm}
could undermine randomness, read keys off buses, and more. We speculate
that, similar to \acp{tpm}, \textbf{most of the systems we studied were
not designed to withstand \ac{pcba} attacks.} \Ac{pcba} security is
expensive to design and implement; if defending against \ac{pcba}
attacks is not a business requirement, why incur this expense?

Attacks on systems that did not even bother to implement basic security
architecture should not be cited as motivation for developing new
countermeasures. These attacks demonstrate that \ac{pcba} defenses
\emph{are needed}, not that existing solutions are inadequate. Although
a few attacks have been proposed that security architecture could not
have helped, for the majority of attacks seen to-date, if only one
defense can be deployed, security architecture is the correct choice.
And this is a happy outcome: as we explained in
Section~\ref{picking_appropriate_defenses}, \ac{pcba} security
architecture has become increasingly accessible because it uses the same
cryptographic primitives as software defenses, which are in high demand.
Today, commercial microcontrollers advertise side channel-hardened
encryption engines, manufacturer-attested cryptographic identities, and
secure provisioning solutions. In contrast, when Microsoft designed the
Xbox~360, it was eventually forced to co-design a secure disc drive
controller with its suppliers after years of exploits enabled by COTS
parts with inadequate security.

Accurately understanding the problem that motivates new work is not just
a pedantic concern. It affects design choices and evaluation methods for
new research. An important question for future work is how precisely a
tamper detection approach must be tailored to a particular kind of
\ac{pcba} attack. For example: does a technique designed to enable a
processor to detect if it has been moved to a different \ac{pcba} as
part of a secret extraction attempt need different sensing methods and
resolution than one designed to detect probing an analog signal? We
expect this question to be answered in the affirmative: for example, a
periodic probe signal designed to characterize trace impedance may
disrupt an analog control circuit more than a digital processor, so
different methods may be required to defend analog vs. digital
circuitry.

Looking ahead, practitioners should mitigate \ac{pcba} tampering via
security architecture based on new commercially-available parts with the
necessary security features, and via careful implementation based on
lessons learned from past systems' mistakes. Meanwhile, researchers
should focus on attacks that cannot be defended by security architecture
and on the specifics of use cases where tamper detection complements
architecture.

Additionally, researchers should focus on the problems highlighted in
Section~\ref{reasoning_about_security}: tamper detection research would
benefit from physical measurements of attacks, countermeasure
performance, and the effects of process/environmental uncertainty.
Several promising \ac{pcba} tamper defense concepts have been proposed,
but we know too little about real-world attacks to assess how effective
these approaches will be in practice. Before we can assess the
likelihood that, e.g., an optical inspection pipeline will detect an
attack, we must learn how similar a ``Trojan'' chip's pixel intensity
histogram could be, when imaged under a microscope, compared with a
legitimate chip. Or: before we can say whether electrical sensors are
capable of detecting a probe wire on a bus, we must understand the
effects of different probes and probing methods on, e.g., impedance and
signal reflections. We need to prove to ourselves that, when reasonable
amounts of noise and process variation are taken into account, it is not
possible for an attacker to implement a malicious chip that looks
exactly like an authentic one under a camera or X-ray, or that it is not
possible to mask signal reflections from probing by tweaking signal
termination impedance or other counter-effects. If such subtle attacks
are possible, we need to be able to reason about their cost and
difficulty so that we are better informed of the ``security return on
investment'' of different approaches. Until we can more rigorously
measure attacks and countermeasures, tamper detection implementors will
need significant research and engineering to validate their
implementation's security.

\section*{Acknowledgments} Thanks to Felix Domke from the Xbox Linux
Project for review of the sections on Xbox and Xbox 360 attacks.

\begin{table*}[ht!]
    \textbf{Appendix:} Summary of \ac{pcba} attack analysis. The third column references the vulnerability classes from Table~\ref{tab:vulnerability_classes}, and the final column marks attacks that, in our assessment, resulted from implementation errors instead of clearly inadequate security architecture. Attacks with more than one class are explained in Section~\ref{sec:survey}.
    \begin{center}
    \begin{tabular}{| m{2.1in} | m{3.25in} | w{c}{.35in} | w{c}{.45in} |}
        \hline
                        		&                  						& 					& \textbf{Impl.}  \\
        \textbf{Attack} 		& \textbf{Summary of vulnerability} 	& \textbf{Class} 	& \textbf{mistake?} \\
        \hline
        ``The Big Hack''~\cite{robertson_big_2018} & Insufficient technical details available & ? & ? \\
        \hline
        ``Modchips of the State''~\cite{hudson_modchips_2018} & No BMC firmware integrity/authenticity checks & 1 &  \\
        \hline
        Emulate admin keystrokes~\cite{hudson_modchips_2018,monta_elkins_nation-state_2019,greenberg_planting_2019} & No protection of physically-accessible admin shell & 5 & \\
        \hline
        Xbox 1BL dump~\cite{huang_keeping_2002,steil_17_2005} 			& Firmware and secrets tx'd in the clear across \ac{pcba} & 2 & \\
        \hline
        Xbox modchips~\cite{steil_17_2005} 			& No effective integrity check & 1 & Y \\
        \hline
        Xbox modchips continue~\cite{andrew_bunnie_huang_hacking_2003,steil_deconstructing_2006} 	& A second failure to implement an integrity check & 1 & Y \\
        \hline
        Xbox ``Visor vulnerability''~\cite{andrew_bunnie_huang_hacking_2003, steil_17_2005} & An important CPU fault did not actually happen & 1 & Y \\
        \hline
        Xbox ``jam code'' interpreter~\cite{andrew_bunnie_huang_hacking_2003, steil_17_2005} & Flawed blacklist of security-compromising operations & 1 & \\
        \hline
        Xbox ``A20 gate'' attack~\cite{steil_17_2005, steil_deconstructing_2006} & Forgotten legacy behavior changes boot vector & 1 & Y \\
        \hline
        Xbox 360 JTAG/SMC~\cite{noauthor_jtag_nodate} & Hypervisor implementation error, no code/data authenticity checks, active JTAG in GPU & 1, 5 & \\
        \hline
        Xbox 360 timing attack~\cite{15432_protecting_2020_part1,copetti-xbox360,noauthor_jtag_nodate} & Calculation time depended on secret value & 1 & Y \\
        \hline
        Xbox 360 RGH~\cite{15432_protecting_2020_part3, noauthor_xbox_2022} & Naive memory comparison, insufficient reset line filtering & 1 & Y \\
        \hline
        Xbox 360 disc drives~\cite{15432_protecting_2020_part1,copetti-xbox360,15432_protecting_2020_part2,chen_guarding_2019,ashcraft_one_2021} & Various design and implementation failures & 1, 2 & Y/N \\
        \hline
        NSA-like debug hacks~\cite{fitzpatrick_nsa_2015,stanislav_mechanics_2014,joe_fitzpatrick_tao_2016} & Presence of active debug infrastructure & Various & \\
        \hline
        NSA-like malicious peripherals~\cite{appelbaum_nsas_2013} & Software vulnerabilities exploitable by untrusted peripherals & 5 & Y \\
        \hline
        NSA drive firmware replacement~\cite{appelbaum_nsas_2013} & MBR parsing vulnerabilities & 4 & Y \\
        \hline
        Evil hard disk controller~\cite{zaddach_implementation_2013} & No authentication of peripheral hardware & 3 & Y \\
        \hline
        Hard drive becomes a microphone~\cite{kwong_hard_2019} & No hard drive firmware integrity checks & 1, 5 & \\
        \hline
        NSA-like ethernet jack trojan~\cite{davis_dock_2013,appelbaum_nsas_2013} & No TLS/PKI & 2, 3, 4 & \\
        \hline
        NSA-like DMA attacks~\cite{appelbaum_nsas_2013, fitzpatrick_tools_2014,inception,frisk_direct_2016} & No limits on peripheral DMA & 4, 5 & \\
        \hline
        NSA-like beacon retroreflectors~\cite{appelbaum_nsas_2013,gbppr} & Tracking device installed on a \ac{pcba} & 6 & \\
        \hline
        NSA microphone retroreflectors~\cite{appelbaum_nsas_2013} & Microphone installed on \ac{pcba} & 6 & \\
        \hline
        NSA-like USB/PS2 retroreflectors~\cite{appelbaum_nsas_2013,michael_ossmann_nsa_2022,wakabayashi_feasibility_2018} & Unprotected CPU $\leftrightarrow$ peripheral signals & 2 & \\
        \hline
        NSA-like video retroreflectors~\cite{appelbaum_nsas_2013,kuhn_electromagnetic_2005} & Analog video signals & 6 & \\
	    \hline
        TPM entropy substitution~\cite{boone_tpm_2018} & No encryption/authentication of inter-IC signals & 4 & \\
        \hline
        TPM secret snooping~\cite{denis_andzakovic_extracting_2019} & Secrets transmitted in the clear across \ac{pcba} & 2 & \\
        \hline
        TPM reset attack~\cite{kauer_oslo_2007,winter_hijackers_2013} & Flawed assumption that TPMs and CPUs reset together & 3 & \\
        \hline
        TPM \acs{drtm} attacks~\cite{winter_hijackers_2013,winter_hijackers_2012} & No encryption/authentication of inter-IC signals & 4 & \\
        \hline
        Malicious laptop dock~\cite{davis_dock_2013} & Unprotected CPU $\leftrightarrow$ peripheral signals & 2, 4, 6 & \\
        \hline
        Early voting machines~\cite{feldman_security_2007, gonggrijp_studying_2007,blaze_report_2017} & No firmware integrity/authenticity checks & 1 & \\
        \hline
        Fingerprint card compromise~\cite{fietkau_swipe_2018, chen2023bruteprint}  & No encryption or authentication of inter-IC signals & 2, 4 & \\
    	\hline
	    Trojan smartphone touchscreen~\cite{shwartz_shattered_2017} & No authentication of security-critical peripheral & 1, 3 & \\
        \hline
	    Glitch attacks using Chip.fail~\cite{thomas_roth_chipfail_2019} & Naive memory comparison, insufficient power/reset filtering & 1 & Y \\
        \hline
	    Altering PCB traces/substrate~\cite{ghosh_how_2015} & Design failure if data or code are affected, open problem if simple digital logic or analog signals are affected & 3, 6 & \\
        \hline
        Add covert channel to existing signal by moving \ac{pcb} traces~\cite{cottais_second_2018} & A trojan trace/signal interferes with the frequency of an oscillator that drives an RF frontend & 6 & \\
	    \hline
        Packet-in-packet attack on TTE~\cite{loveless_pcspoof_2023} & No encryption/authentication of digital traffic & 2, 4 & \\
        \hline
        Tampering analog control~\cite{ghosh_how_2015} & Additing discrete components causes analog control failure & 6 & \\
        \hline
        Destructive high voltage~\cite{noauthor_usbkill_nodate} & Destroy a component by exposing to over-voltage & 6 & \\
        \hline
        Trace breakage~\cite{ghosh_how_2015, mcguire_pcb_2019} & Thin traces to induce electromigration & 6 & \\
        \hline
    \end{tabular}
    \end{center}
\end{table*}


\begin{thebibliography}{100}

\bibitem{15432_protecting_2020_part1}
15432.
\newblock Protecting and hacking the {Xbox} 360 (part 1).
\newblock discourse.world, March 2020.

\bibitem{15432_protecting_2020_part2}
15432.
\newblock Protecting and hacking the {Xbox} 360 (part 2).
\newblock discourse.world, April 2020.

\bibitem{15432_protecting_2020_part3}
15432.
\newblock Protecting and hacking the {Xbox} 360 (part 3).
\newblock discourse.world, May 2020.

\bibitem{amd_amd_nodate}
AMD.
\newblock {AMD} secure encrypted virtualization ({SEV}).
\newblock Developer Central.

\bibitem{anderson_security_2020}
Ross Anderson.
\newblock {\em Security Engineering: A Guide to Building Dependable Distributed
  Systems}.
\newblock John Wiley, Newark, New Jersey, 2020.

\bibitem{denis_andzakovic_extracting_2019}
Denis Andzakovic.
\newblock Extracting {Bitlocker} keys from a {TPM}.
\newblock Pulse Security, March 2019.

\bibitem{appelbaum_nsas_2013}
Jacob Appelbaum, Judith Horchert, Ole Reissmann, Marcel Rosenbach, J{\"o}rg
  Schindler, and Christian St{\"o}cker.
\newblock {NSA}'s secret toolbox: Unit offers spy gadgets for every need.
\newblock {\em Der Spiegel}, December 2013.

\bibitem{apple-iommu}
{Apple}.
\newblock Direct memory access protections for {Mac} computers.
\newblock Apple Platform Security documentation, 2021.

\bibitem{ashcraft_one_2021}
Brian Ashcraft.
\newblock One of the wildest console hacks ever, 2021.

\bibitem{bergman_battelle_2016}
Thomas~D. Bergman, Cyber~Program Manager, and Katie~T. Liszewski.
\newblock Battelle barricade: {A} nondestructive electronic component
  authentication and counterfeit detection technology.
\newblock In {\em 2016 {IEEE} {THS}}, pages 1--6, 2016.

\bibitem{bhattacharyay_vipr-pcb_2022}
Aritra Bhattacharyay, Prabuddha Chakraborty, Jonathan Cruz, and Swarup Bhunia.
\newblock {VIPR}-{PCB}: {A} {Machine} {Learning} {Based} {Golden}-{Free} {PCB}
  {Assurance} {Framework}.
\newblock In {\em Proceedings of the 59th {ACM}/{IEEE} {Design} {Automation}
  {Conference}}, pages 793--798, 2022.

\bibitem{bhattacharyay_automated_2022}
Aritra Bhattacharyay, Shuo Yang, Jonathan Cruz, Prabuddha Chakraborty, Swarup
  Bhunia, and Tamzidul Hoque.
\newblock An {Automated} {Framework} for {Board}-level {Trojan} {Benchmarking}.
\newblock {\em IEEE Transactions on Computer-Aided Design of Integrated
  Circuits and Systems}, pages 1--1, 2022.

\bibitem{bhunia_editorial_2017}
Swarup Bhunia and Mark Tehranipoor.
\newblock Editorial for the {Introductory} {Issue} of the {Journal} of
  {Hardware} and {Systems} {Security} ({HaSS}).
\newblock {\em Journal of Hardware and Systems Security}, pages 1--2, March
  2017.

\bibitem{blaze_report_2017}
Matt Blaze, Jake Braun, Harri Hursti, Joseph Lorenzo~Hall, Margaret MacAlpine,
  and Jeff Moss.
\newblock Report on {Cyber} {Vulnerabilities} in {U}.{S}. {Election}
  {Equipment}, {Databases}, and {Infrastructure}, September 2017.

\bibitem{ruxcon-hit-by-bus}
Adam Boileau.
\newblock Hit by a bus: physical access attacks with {Firewire}.
\newblock In {\em Ruxcon}, 2006.

\bibitem{boone_tpm_2018}
Jeremy Boone.
\newblock {TPM} {Genie}: {Interposer} {Attacks} {Against} the {Trusted}
  {Platform} {Module} {Serial} {Bus}, 2018.
\newblock Pages: 1-23.

\bibitem{9707715}
Ulbert~J. Botero, Fatemeh Ganji, Damon~L. Woodard, and Domenic Forte.
\newblock Automated trace and copper plane extraction of x-ray tomography
  imaged pcbs.
\newblock In {\em IEEE PAINE Proceedings}, pages 1--8, 2021.

\bibitem{botero_automated_2021}
Ulbert~J. Botero, Fatemeh Ganji, Damon~L. Woodard, and Domenic Forte.
\newblock Automated {Trace} and {Copper} {Plane} {Extraction} of {X}-ray
  {Tomography} {Imaged} {PCBs}.
\newblock In {\em 2021 {IEEE} {Physical} {Assurance} and {Inspection} of
  {Electronics} ({PAINE})}, pages 1--8, November 2021.

\bibitem{botero_hardware_2021}
Ulbert~J. Botero, Ronald Wilson, Hangwei Lu, Mir~Tanjidur Rahman, Mukhil~A.
  Mallaiyan, Fatemeh Ganji, Navid Asadizanjani, Mark~M. Tehranipoor, Damon~L.
  Woodard, and Domenic Forte.
\newblock Hardware {Trust} and {Assurance} through {Reverse} {Engineering}: {A}
  {Tutorial} and {Outlook} from {Image} {Analysis} and {Machine} {Learning}
  {Perspectives}.
\newblock {\em J. Emerg. Technol. Comput. Syst.}, 17(4), June 2021.

\bibitem{sergey_bratus_invited_2017}
Sergey Bratus and Anna Shubina.
\newblock Overlooked foundations: Exploits as experiments and constructive
  proofs in the science-of-security.
\newblock In {\em Proceedings of the 10th USENIX Conference on Cyber Security
  Experimentation and Test Proceedings of the 10th USENIX Conference on Cyber
  Security Experimentation and Test}, Vancouver, BC, August 2017. USENIX.

\bibitem{chen_guarding_2019}
Tony Chen.
\newblock Guarding {Against} {Physical} {Attacks}: {The} {Xbox} {One} {Story}.
\newblock Presented at Platform Security Summit, October 2019.

\bibitem{chen2023bruteprint}
Yu~Chen and Yiling He.
\newblock Bruteprint: Expose smartphone fingerprint authentication to
  brute-force attack.
\newblock arXiv 2305.10791, 2023.

\bibitem{component_detection}
Deruo Cheng, Jingyang Dai, Yee-Yang Tee, Yiqiong Shi, and Bah-Hwee Gwee.
\newblock {PCB} surface component detection with computer vision assisted label
  generation.
\newblock In {\em {IEEE} {IPFA} Proceedings}, 2024.

\bibitem{cobb_intrinsic_2012}
William~E. Cobb, Eric~D. Laspe, Rusty~O. Baldwin, Michael~A. Temple, and
  Yong~C. Kim.
\newblock Intrinsic physical-layer authentication of integrated circuits.
\newblock {\em IEEE Transactions on Information Forensics and Security},
  7(1):14--24, 2012.

\bibitem{copetti_xbox_2020}
Rodrigo Copetti.
\newblock Xbox architecture - a practical analysis, 2020.

\bibitem{copetti-xbox360}
Rodrigo Copetti.
\newblock Xbox 360 architecture - a practical analysis, 2022.

\bibitem{cottais_second_2018}
Emmanuel Cottais, Jose~Lopes Esteves, and Chaouki Kasmi.
\newblock Second {Order} {Soft}-{TEMPEST} in {RF} {Front}-{Ends}: {Design} and
  {Detection} of {Polyglot} {Modulations}.
\newblock In {\em 2018 {International} {Symposium} on {Electromagnetic}
  {Compatibility} ({EMC} {EUROPE})}, pages 166--171, 2018.

\bibitem{10647049}
Patrick Craig, Antika Roy, Nitin Varshney, and Navid Asadizanjani.
\newblock Advancing {PCB} assurance towards netlist extraction with the
  integration of {X-Ray} imaging and semi-supervised learning techniques.
\newblock In {\em 2024 IEEE Research and Applications of Photonics in Defense
  Conference (RAPID)}, pages 1--2, 2024.

\bibitem{davis_dock_2013}
Andy Davis.
\newblock To dock or not to dock, that is the question: using laptop docking
  stations as hardware-based attack platforms.
\newblock In {\em BlackHat Europe}, Amsterdam, Netherlands, 2013.

\bibitem{dhanuskodi_counterfoil_2020}
Siva~Nishok Dhanuskodi, Xiang Li, and Daniel Holcomb.
\newblock {COUNTERFOIL}: Verifying provenanc of integrated circuits using
  intrinsic package fingerprints and inexpensive cameras.
\newblock In {\em 29th {USENIX} Security Symposium}, pages 1255--1272, August
  2020.

\bibitem{edwards_authenticating_2017}
Nathan Edwards, Jason Hamlet, and Mitchell~T. Martin.
\newblock Authenticating a printed circuit board.
\newblock US patent 10594492B1, 2017.

\bibitem{monta_elkins_nation-state_2019}
Monta Elkins.
\newblock Nation-state supply chain attacks for dummies and you too, 2019.

\bibitem{fail0verflow_console_nodate}
fail0verflow.
\newblock Console {Hacking} 2010.
\newblock Presented at 27C3.

\bibitem{farmer_sold_2014}
Dan Farmer.
\newblock Sold down the river, 2014.

\bibitem{feldman_security_2007}
Ariel~J. Feldman, J.~Alex Halderman, and Edward~W. Felten.
\newblock Security analysis of the {Diebold} {AccuVote}-{TS} voting machine.
\newblock In {\em Proceedings of the {USENIX} {Workshop} on {Accurate}
  {Electronic} {Voting} {Technology}}, page~2, 2007.

\bibitem{fietkau_swipe_2018}
Julian Fietkau, {Starbug}, and Jean-Pierre Seifert.
\newblock Swipe your fingerprints! {How} biometric authentication simplifies
  payment, access and identity fraud.
\newblock In {\em {USENIX} {WOOT}}, August 2018.

\bibitem{inception}
{Fist0urs}, {maxgrim}, {carmaa}, and {rexploit}.
\newblock Inception password unlocking payload: inception/modules/unlock.py.
\newblock GitHub, September 2017.

\bibitem{joe_fitzpatrick_tao_2016}
Joe Fitzpatrick.
\newblock The {Tao} of {Hardware}, {The} {Te} of {Implants}.
\newblock Blackhat USA, 2016.

\bibitem{fitzpatrick_nsa_2015}
Joe FitzPatrick and Matt King.
\newblock {NSA} {Playset}: {JTAG} implants.
\newblock In {\em Defcon 23}, 2015.

\bibitem{fitzpatrick_tools_2014}
Joe FitzPatrick and Mike Ryan.
\newblock Tools of the {NSA} playset.
\newblock In {\em Ruxcon}, 2014.

\bibitem{free60_boot_nodate}
Boot process.
\newblock Free60 Wiki Archive.

\bibitem{free60_bootloaders_nodate}
Bootloaders.
\newblock Free60 Wiki Archive.

\bibitem{free60_nand}
Free60.
\newblock Nand.
\newblock Free60 Wiki Archive.

\bibitem{free60_fusesets}
Free60.
\newblock Understanding the {Xbox} 360's fusesets.
\newblock Free60 Wiki Archive.

\bibitem{noauthor_jtag_nodate}
The {JTAG}/{SMC} hack.
\newblock Technical description of JTAG/SMC hack, bundled with the utility as a
  ReadMe. The Free60 version contains images that are not in the ReadMe., Nov
  2009.

\bibitem{noauthor_xbox_2022}
The {Xbox} 360 reset glitch hack.
\newblock Free60 Wiki Archive, February 2022.

\bibitem{frisk_direct_2016}
Ulf Frisk.
\newblock Direct {Memory} {Attack} the {Kernel}.
\newblock In {\em {DEF} {CON} 24}, 2016.

\bibitem{ghosh_recycled_2019}
Pallabi Ghosh and Rajat~Subhra Chakraborty.
\newblock Recycled and remarked counterfeit integrated circuit detection by
  image-processing-based package texture and indent analysis.
\newblock {\em IEEE Transactions on Industrial Informatics}, 15(4):1966--1974,
  2019.

\bibitem{ghosh_how_2015}
Swaroop Ghosh, Abhishek Basak, and Swarup Bhunia.
\newblock How secure are printed circuit boards against trojan attacks?
\newblock {\em IEEE Design Test}, 32(2):7--16, 2015.

\bibitem{gonggrijp_studying_2007}
Rop Gonggrijp and Willem-Jan Hengeveld.
\newblock Studying the {Nedap}/{Groenendaal} {ES3B} voting computer: A computer
  security perspective.
\newblock In {\em Proceedings of the {USENIX} {Workshop} on {Accurate}
  {Electronic} {Voting} {Technology}}, page~1, 2007.

\bibitem{gbppr}
{Green Bay Professional Packet Radio}.
\newblock {GBPPR2}.
\newblock {YouTube} channel.

\bibitem{greenberg_planting_2019}
Andy Greenberg.
\newblock Planting tiny spy chips in hardware can cost as little as \$200.
\newblock {\em WIRED}, October 2019.

\bibitem{glenn_greenwald_how_2014}
Glenn Greenwald.
\newblock How the {NSA} tampers with {US}-made internet routers.
\newblock {\em The Guardian}, May 2014.

\bibitem{guo_mpa_2017}
Zimu Guo, Xiaolin Xu, Mark~M. Tehranipoor, and Domenic Forte.
\newblock {MPA}: {Model}-assisted {PCB} attestation via board-level {RO} and
  temperature compensation.
\newblock In {\em 2017 {Asian} {Hardware} {Oriented} {Security} and {Trust}
  {Symposium} ({AsianHOST})}, pages 25--30, 2017.

\bibitem{guo_eop_2019}
Zimu Guo, Xiaolin Xu, Mark~M. Tehranipoor, and Domenic Forte.
\newblock {EOP}: An encryption-obfuscation solution for protecting {PCBs}
  against tampering and reverse engineering.
\newblock {\em arXiv: 1904.09516}, 2019.

\bibitem{laser_mark_analysis}
Jacob Harrison, Nathan Jessurun, Raphael R.~Dos Santos, Shajib Ghosh, Navid
  Asadi, and Mark Tehranipoor.
\newblock Analysis of etcher configuration on part marking characteristics for
  counterfeit identification.
\newblock In {\em {IEEE} {IPFA} Proceedings}, 2024.

\bibitem{heasman_implementing_2007}
John Heasman.
\newblock Implementing and detecting a {PCI} rootkit, 2007.

\bibitem{hernandez_firmusb_2017}
Grant Hernandez, Farhaan Fowze, Dave~(Jing) Tian, Tuba Yavuz, and Kevin~R.B.
  Butler.
\newblock {FirmUSB}: Vetting {USB} device firmware using domain informed
  symbolic execution.
\newblock In {\em Proceedings of the 2017 {ACM} {SIGSAC} {Conference} on
  {Computer} and {Communications} {Security}}, 2017.

\bibitem{huang_keeping_2002}
Andrew Huang.
\newblock Keeping secrets in hardware: the {Microsoft} {XBox} case study, May
  2002.
\newblock AI Memo 2002-08.

\bibitem{andrew_bunnie_huang_hacking_2003}
Andrew Huang.
\newblock {\em Hacking the {Xbox}: an introduction to reverse engineering}.
\newblock No Starch Press, San Francisco, CA, 2003.

\bibitem{hudson_modchips_2018}
Trammel Hudson.
\newblock Modchips of the state.
\newblock in 35C3, 2018.

\bibitem{immler_b-trepid_2018}
Vincent Immler, Johannes Obermaier, Martin K{\"o}nig, Matthias Hiller, and
  Georg Sig.
\newblock B-{TREPID}: {Batteryless} tamper-resistant envelope with a {PUF} and
  integrity detection.
\newblock In {\em 2018 {IEEE} {HOST}}, pages 49--56, April 2018.

\bibitem{noauthor_xbox_2007}
Xbox 360 timing attack.
\newblock IVC Wiki, December 2007.

\bibitem{10.1145/3588032}
Nathan Jessurun, Olivia~P. Dizon-Paradis, Jacob Harrison, Shajib Ghosh, Mark~M.
  Tehranipoor, Damon~L. Woodard, and Navid Asadizanjani.
\newblock {FPIC}: A novel semantic dataset for optical {PCB} assurance.
\newblock {\em J. Emerg. Technol. Comput. Syst.}, 19(2), September 2023.

\bibitem{pinpoint}
Nathan Jessurun, Jacob Harrison, Mark~M. Tehranipoor, and Navid Asadizanjani.
\newblock Pinpoint: An smd pin localization method.
\newblock In {\em {IEEE} {IPFA} Proceedings}, 2022.

\bibitem{jessurun_component_2020}
Nathan Jessurun, Olivia Paradis, Alexandra Roberts, and Navid Asadizanjani.
\newblock Component detection and evaluation framework ({CDEF}): {A} semantic
  annotation tool.
\newblock {\em Microscopy and Microanalysis}, 26(S2):1470--1474, August 2020.

\bibitem{karri_fuzzingcontrolled_2021}
R~Karri, F~Khorrami, and P~Krishnamurthy.
\newblock Fuzzing/controlled excitation and multi-modal sensor
  monitoring/fusion for hardware-firmware-software integrity verification.
\newblock Technical report, NYU Tandon School of Engineering, July 2021.

\bibitem{kauer_oslo_2007}
Bernhard Kauer.
\newblock {OSLO}: {Improving} the {Security} of {Trusted} {Computing}.
\newblock In {\em Proceedings of 16th {USENIX} {Security} {Symposium} on
  {USENIX} {Security} {Symposium}}, {SS}'07, USA, 2007. USENIX Association.
\newblock event-place: Boston, MA.

\bibitem{10.1145/3606948}
David~Selasi Koblah, Ulbert~J. Botero, Sean~P. Costello, Olivia~P.
  Dizon-Paradis, Fatemeh Ganji, Damon~L. Woodard, and Domenic Forte.
\newblock A fast object detection-based framework for via modeling on {PCB}
  {X-Ray} {CT} images.
\newblock {\em ACM J. Emerg. Technol. Comput. Syst.}, 19(4), oct 2023.

\bibitem{kuhn_electromagnetic_2005}
Markus~G. Kuhn.
\newblock Electromagnetic eavesdropping risks of flat-panel displays.
\newblock In {\em Privacy {Enhancing} {Technologies}}, 2005.

\bibitem{kumar_devfing_2021}
Vijay Kumar and Kolin Paul.
\newblock {DevFing}: {Robust} {LCR} based device fingerprinting.
\newblock In {\em 10th {MECO} Proceedings}, pages 1--6, June 2021.

\bibitem{kwong_hard_2019}
Andrew Kwong, Wenyuan Xu, and Kevin Fu.
\newblock Hard {Drive} of {Hearing}: {Disks} that {Eavesdrop} with a
  {Synthesized} {Microphone}.
\newblock In {\em 2019 {IEEE} {Symposium} on {Security} and {Privacy} ({SP})},
  pages 905--919, 2019.

\bibitem{lawson_tpm_2007}
Nate Lawson.
\newblock {TPM} hardware attacks (part 2).
\newblock rdist, July 2007.

\bibitem{darpa_shield}
Serge Leef.
\newblock Supply chain hardware integrity for electronics defense ({SHIELD}).
\newblock {DARPA} Software and Supply Chain Assurance Winter Forum, December
  2018.

\bibitem{loveless_pcspoof_2023}
A.~Loveless, L.~Phan, R.~Dreslinski, and B.~Kasikci.
\newblock {PCspooF}: Compromising the safety of time-triggered ethernet.
\newblock In {\em {IEEE} {Symposium} on {Security} and {Privacy}}, pages
  572--587, May 2023.

\bibitem{MAILLARD2023104904}
Julien Maillard, Thomas Hiscock, Maxime Lecomte, and Christophe Clavier.
\newblock Side-channel disassembly on a system-on-chip: A practical feasibility
  study.
\newblock {\em Microprocessors and Microsystems}, 101, 2023.

\bibitem{martin_amds_2019}
Dylan Martin.
\newblock {AMD}'s {Xbox}, {PlayStation} work led to a big security feature in
  {EPYC}.
\newblock {\em CRN}, August 2019.

\bibitem{mcguire_pcb_2019}
Matthew McGuire, Umit Ogras, and Sule Ozev.
\newblock {PCB} hardware trojans: Attack modes and detection strategies.
\newblock In {\em 2019 {IEEE} 37th {VLSI} {Test} {Symposium} ({VTS})}, pages
  1--6, 2019.

\bibitem{10.1145/3401980}
Dhwani Mehta, Hangwei Lu, Olivia~P. Paradis, Mukhil~M.S. Azhagan, M.~Tanjidur
  Rahman, Yousef Iskander, Praveen Chawla, Damon~L. Woodard, Mark Tehranipoor,
  and Navid Asadizanjani.
\newblock The big hack explained: Detection and prevention of {PCB} supply
  chain implants.
\newblock {\em J. Emerg. Technol. Comput. Syst.}, 16(4), August 2020.

\bibitem{cryptoeprint:2022/924}
Dhwani Mehta, John True, Olivia~P. Dizon-Paradis, Nathan Jessurun, Damon~L.
  Woodard, Navid Asadizanjani, and Mark Tehranipoor.
\newblock {FICS} {PCB} x-ray: A dataset for automated printed circuit board
  inter-layers inspection.
\newblock Cryptology {ePrint} Archive, Paper 2022/924, 2022.

\bibitem{virtualization-security}
Microsoft.
\newblock Virtualization-based security ({VBS}).
\newblock Windows Hardware Developer documentation, 2017.

\bibitem{ImpedanceVerif_2022}
Tahoura Mosavirik, Patrick Schaumont, and Shahin Tajik.
\newblock {ImpedanceVerif}: {On}-chip impedance sensing for system-level
  tampering detection.
\newblock {\em {IACR} {TCHES}}, 2023(1):301--325, Nov. 2022.

\bibitem{nishizawa_capacitance_2018}
Makoto Nishizawa, Kento Hasegawa, and Nozomu Togawa.
\newblock Capacitance {Measurement} of {Running} {Hardware} {Devices} and its
  {Application} to {Malicious} {Modification} {Detection}.
\newblock In {\em 2018 {IEEE} {Asia} {Pacific} {Conference} on {Circuits} and
  {Systems} ({APCCAS})}, pages 362--365, 2018.

\bibitem{nsa_developing_2012}
NSA.
\newblock Developing a blueprint for a science of cybersecurity.
\newblock {\em The Next Wave}, 19(2), 2012.

\bibitem{obermaier_shedding_2017}
Johannes Obermaier and Stefan Tatschner.
\newblock Shedding too much {Light} on a {Microcontroller}'s {Firmware}
  {Protection}.
\newblock In {\em 11th {USENIX} {Workshop} on {Offensive} {Technologies}},
  August 2017.

\bibitem{oksman_method_2020}
Aapo Oksman.
\newblock {\em A Method for Detecting {DRAM} Bus Tampering}.
\newblock PhD thesis, Aalto University, 2020.

\bibitem{michael_ossmann_nsa_2022}
Michael Ossmann.
\newblock The {NSA} {Playset}: {RF} retroreflectors.
\newblock In {\em Defcon 22}, Las Vegas, NV, USA, 2014.

\bibitem{paley_active_2016}
Steven Paley, Tamzidul Hoque, and Swarup Bhunia.
\newblock Active protection against {PCB} physical tampering.
\newblock In {\em 2016 17th {International} {Symposium} on {Quality}
  {Electronic} {Design} ({ISQED})}, pages 356--361, 2016.

\bibitem{paul_silverin_2021}
Shubhra~Deb Paul and Swarup Bhunia.
\newblock {SILVerIn}: Systematic integrity verification of printed circuit
  board using {JTAG} infrastructure.
\newblock {\em J. Emerg. Technol. Comput. Syst.}, 17(3), June 2021.

\bibitem{pearce_detecting_2022}
Hammond Pearce, Virinchi~Roy Surabhi, Prashanth Krishnamurthy, Joshua Trujillo,
  Ramesh Karri, and Farshad Khorrami.
\newblock Detecting hardware trojans in {PCBs} using side channel loopbacks.
\newblock {\em IEEE Trans. Very Large Scale Integr. Syst.}, 30(7):926--937,
  July 2022.

\bibitem{piliposyan_computer_2022}
Gor Piliposyan and Saqib Khursheed.
\newblock Computer {Vision} for {Hardware} {Trojan} {Detection} on a {PCB}
  {Using} {Siamese} {Neural} {Network}.
\newblock In {\em 2022 {IEEE} {Physical} {Assurance} and {Inspection} of
  {Electronics} ({PAINE})}, pages 1--7, 2022.

\bibitem{piliposyan_hardware_2020}
Gor Piliposyan, Saqib Khursheed, and Daniele Rossi.
\newblock Hardware trojan detection on a {PCB} through differential power
  monitoring.
\newblock {\em IEEE Transactions on Emerging Topics in Computing}, 2020.

\bibitem{robertson_big_2018}
J.~Robertson and M.~Riley.
\newblock The big hack: How {China} used a tiny chip to infiltrate {U}.{S}.
  companies.
\newblock Bloomberg Businessweek, 2018.

\bibitem{rosenfeld_attacks_2010}
Kurt Rosenfeld and Ramesh Karri.
\newblock Attacks and defenses for {JTAG}.
\newblock {\em IEEE Design Test of Computers}, 27(1):36--47, 2010.
\newblock Number: 1.

\bibitem{thomas_roth_chipfail_2019}
Thomas Roth, Josh Datko, and Dmitry Nedospasov.
\newblock Chip.fail, 2019.

\bibitem{russ_three_2020}
Samuel Russ and Jacob Gatlin.
\newblock Three {Ways} to {Hack} a {Printed} {Circuit} {Board}.
\newblock {\em IEEE Spectrum}, August 2020.
\newblock Publication Title: IEEE Spectrum.

\bibitem{scattering_param_2023}
Maryam~Saadat Safa, Tahoura Mosavirik, and Shahin Tajik.
\newblock Counterfeit chip detection using scattering parameter analysis.
\newblock In {\em 26th {DDECS}}, pages 99--104, 2023.

\bibitem{safa2024parasiticcircusonfeasibilitygolden}
Maryam~Saadat Safa, Patrick Schaumont, and Shahin Tajik.
\newblock Parasitic circus: On the feasibility of golden free {PCB}
  verification.
\newblock In {\em {IEEE} {IPFA} Proceedings}, 2024.

\bibitem{salihun_nsa_2014_pt2}
Darmawan Salihun.
\newblock {NSA} backdoor part 2, {BULLDOZER}: And, learn how to {DIY} a {NSA}
  hardware implant.
\newblock Infosec, February 2014.

\bibitem{salihun_nsa_2014_pt1}
Darmawan Salihun.
\newblock {NSA} {BIOS} {Backdoor} a.k.a. {God} {Mode} {Malware} {Part} 1:
  {DEITYBOUNCE}.
\newblock Infosec, January 2014.

\bibitem{10.31399/asm.cp.istfa2021p0012}
Mukhil Azhagan~Mallaiyan Sathiaseelan, Olivia~P. Paradis, Rajat Rai,
  Suryaprakash~Vasudev Pandurangi, Manoj~Yasaswi Vutukuru, Shayan Taheri, and
  Navid Asadizanjani.
\newblock Logo classification and data augmentation techniques for {PCB}
  assurance and counterfeit detection.
\newblock In {\em ISTFA 47 Proceedings}, oct 2021.

\bibitem{schaumont_hide_but_cant_verify}
Patrick Schaumont.
\newblock You can hide but you can't verify: On side-channel countermeasure
  verification.
\newblock In {\em Workshop on SSH-SoC: Safety and Security in Heterogeneous
  Open System-on-Chip Platforms}, 2023.

\bibitem{shwartz_shattered_2017}
Omer Shwartz, Amir Cohen, Asaf Shabtai, and Yossi Oren.
\newblock Shattered trust: When replacement smartphone components attack.
\newblock In {\em 11th {USENIX} Workshop on Offensive Technologies}. USENIX
  Association, August 2017.

\bibitem{dominic_spill_nsa_2015}
Dominic Spill.
\newblock {NSA} {Playset}: {USB} {Tools}.
\newblock Shmoocon, 2015.

\bibitem{st-sfi}
{ST Microelectronics}.
\newblock Introduction to secure firmware install ({SFI}) for {STM32} {MCUs}.
\newblock AN4992, 2023.

\bibitem{stanislav_mechanics_2014}
{Stanislav}.
\newblock Mechanics of {FLUXBABBITT}.
\newblock Loper OS, January 2014.

\bibitem{steil_17_2005}
Michaael Steil.
\newblock 17 mistakes {Microsoft} made in the {Xbox} security system, 2005.

\bibitem{steil_deconstructing_2006}
Michaael Steil.
\newblock Deconstructing the {Xbox} `security system', December 2006.

\bibitem{steil_xbox_2008}
Michaael Steil and Felix Domke.
\newblock The {Xbox} 360 security system and its weaknesses, 2008.

\bibitem{tcg_tpm_2011}
TCG.
\newblock {TPM} main specification: Part 1 design principles, March 2011.
\newblock v1.2.

\bibitem{true_review_2021}
John True, Chengjie Xi, Nathan Jessurun, Kiarash Ahi, and Navid Asadizanjani.
\newblock Review of {THz}-based semiconductor assurance.
\newblock {\em Optical Engineering}, 60(6):1 -- 52, 2021.

\bibitem{tcg-drtm}
{Trusted Computing Group}.
\newblock {TCG} {D}-{RTM} architecture.
\newblock Specification, June 2013.

\bibitem{noauthor_usbkill_nodate}
{USBKill}.

\bibitem{wakabayashi_feasibility_2018}
Satohiro Wakabayashi, Seita Maruyama, Tatsuya Mori, Shigeki Goto, Masahiro
  Kinugawa, Yu-ichi Hayashi, and Michael Smith.
\newblock A {Feasibility} {Study} of {Radio}-frequency {Retroreflector}
  {Attack}.
\newblock In {\em 12th {USENIX} {Workshop} on {Offensive} {Technologies}},
  Baltimore, MD, August 2018. USENIX Association.

\bibitem{wang_system-level_2019}
Xiaoxiao Wang, Yueying Han, and Mark Tehranipoor.
\newblock System-{Level} {Counterfeit} {Detection} {Using} {On}-{Chip} {Ring}
  {Oscillator} {Array}.
\newblock {\em IEEE Transactions on Very Large Scale Integration (VLSI)
  Systems}, 27(12):2884--2896, December 2019.
\newblock Number: 12.

\bibitem{winter_hijackers_2012}
Johannes Winter and Kurt Dietrich.
\newblock A hijacker's guide to the {LPC} bus.
\newblock In {\em Public {Key} {Infrastructures}, {Services} and
  {Applications}}, pages 176--193, 2012.

\bibitem{winter_hijackers_2013}
Johannes Winter and Kurt Dietrich.
\newblock A hijacker's guide to communication interfaces of the trusted
  platform module.
\newblock {\em Computers \& Mathematics with Applications}, 65(5):748--761,
  2013.

\bibitem{noauthor_xbox_nodate}
Xbox game disc.
\newblock XboxDevWiki.

\bibitem{xu_runtime_2021}
Zhenyu Xu, Thomas Mauldin, Qing Yang, and Tao Wei.
\newblock Runtime {Detection} of {Probing}/{Tampering} on {Interconnecting}
  {Buses}.
\newblock In {\em 2021 {IEEE} 29th {Annual} {International} {Symposium} on
  {Field}-{Programmable} {Custom} {Computing} {Machines} ({FCCM})}, pages
  247--251, 2021.

\bibitem{xu_bus_2020}
Zhenyu Xu, Thomas Mauldin, Zheyi Yao, Shuyi Pei, Tao Wei, and Qing Yang.
\newblock A bus authentication and anti-probing architecture extending hardware
  trusted computing base off {CPU} chips and beyond.
\newblock In {\em 2020 {ACM}/{IEEE} 47th {Annual} {International} {Symposium}
  on {Computer} {Architecture} ({ISCA})}, pages 749--761, 2020.

\bibitem{zaddach_implementation_2013}
Jonas Zaddach, Anil Kurmus, Davide Balzarotti, Erik-Oliver Blass, Aur{\'e}lien
  Francillon, Travis Goodspeed, Moitrayee Gupta, and Ioannis Koltsidas.
\newblock Implementation and implications of a stealth hard-drive backdoor.
\newblock In {\em Proceedings of the 29th Annual Computer Security Applications
  Conference}, pages 279--288, 2013.

\bibitem{zhang_robust_2015}
Fengchao Zhang, Andrew Hennessy, and Swarup Bhunia.
\newblock Robust counterfeit {PCB} detection exploiting intrinsic trace
  impedance variations.
\newblock In {\em 33rd {IEEE} {VTS}}, pages 1--6, 2015.

\bibitem{zhang_database-free_2021}
Fengchao Zhang, Shubhra~Deb Paul, Patanjali Slpsk, Amit~Ranjan Trivedi, and
  Swarup Bhunia.
\newblock On {Database}-{Free} {Authentication} of {Microelectronic}
  {Components}.
\newblock {\em IEEE Transactions on Very Large Scale Integration (VLSI)
  Systems}, 29(1), 2021.

\bibitem{zheng_design_2017}
Xiaomin Zheng, Yuejun Zhang, Jiaweng Zhang, and Wenqi Hu.
\newblock Design impedance mismatch physical unclonable functions for {IoT}
  security.
\newblock {\em Active and Passive Electronic Components}, January 2017.

\bibitem{zhu_pcbench_2021}
Huifeng Zhu, Xiaolong Guo, Yier Jin, and Xuan Zhang.
\newblock {PCBench}: {Benchmarking} of {Board}-{Level} {Hardware} {Attacks} and
  {Trojans}.
\newblock In {\em 2021 26th {Asia} and {South} {Pacific} {Design} {Automation}
  {Conference} ({ASP}-{DAC})}, pages 396--401, 2021.

\bibitem{10153638}
Huifeng Zhu, Haoqi Shan, Dean Sullivan, Xiaolong Guo, Yier Jin, and Xuan Zhang.
\newblock {PDNPulse}: Sensing {PCB} anomaly with the intrinsic power delivery
  network.
\newblock {\em IEEE Transactions on Information Forensics and Security},
  18:3590--3605, 2023.

\end{thebibliography}
\end{document}